\documentstyle[amssymb,aps,preprint,epsf,psfig]{revtex}

\begin{document}
\title{Mirages and many-body effects in quantum corrals}
\author{A. A. Aligia and A.M. Lobos}
\address{Comisi\'{o}n Nacional de Energ\'{\i }a At\'{o}mica,\\
Centro At\'{o}mico Bariloche and Instituto Balseiro, 8400 Bariloche,\\
Argentina }
\maketitle

\begin{abstract}
In the experiment of the quantum mirage, confinement of surface states in an
elliptical corral has been used to project the Kondo effect from one focus
where a magnetic impurity was placed, to the other empty focus. The signature of
the Kondo effect is seen as a Fano antiresonance in scanning tunneling
spectroscopy. This experiment combines the many-body physics of the Kondo
effect with the subtle effects of confinement. In this work we review the
essential physics of the quantum mirage experiment, and present new
calculations involving other geometries and more than one impurity in the
corral, which should be relevant for other experiments that are being made,
and to discern the relative importance of the hybridization of the impurity
with surface ($V_{s}$) and bulk ($V_{b}$) states. The intensity of the
mirage imposes a lower bound to $V_{s}/V_{b}$ which we estimate. Our
emphasis is on the main physical ingredients of the phenomenon and the
many-body aspects, like the dependence of the observed differential
conductance with geometry, which cannot be calculated with alternative
one-body theories. The system is described with an Anderson impurity model
solved using complementary approaches: perturbation theory in the Coulomb
repulsion $U$, slave bosons in mean field and exact diagonalization plus
embedding.
\end{abstract}


\section{Introduction}

The study of many-body phenomena in nanoscale systems is attracting a lot of
attention in recent years. Progress in nanotechnology made it possible to
construct nanodevices such as quantum dots (QD's) which act as ideal
one-impurity systems in which the Kondo physics is clearly displayed \cite
{gold1,cron,gold2,wiel}. The spectral density of localized electrons of a
magnetic impurity in a metallic host, described by the impurity Anderson model 
\cite{pwa}, is known to display a resonance near the Fermi energy in the
localized (or Kondo) regime \cite{hewson}. The conductance through a QD is
proportional to this density and calculations of the latter using the Wilson
renormalization group leads to a good agreement with experiment \cite
{izum,costi}. In contrast to the density of localized electrons, the density
of conduction electrons coupled to the former shows a dip or Fano
antiresonance \cite{fano} (see section 3). Using the fact that the tip of
the scanning tunneling microscope (STM) captures essentially conduction
electrons, Fano line shapes have been observed using scanning tunneling
spectroscopy (STS) for different cases of magnetic impurities on metal
surfaces \cite{li,mad,man,man2,jam,mad2,naga,knorr,schn,wahl}. They should also
manifest in the conductance through quantum wires side-coupled to QD's \cite
{bul,kang,torio,pro,torio2,lady}. The impurity Anderson model also describes
the physics of the conductance through arrays of QD's weakly coupled to
conducting leads \cite{ihm}.

A quantum corral is an area of about 40 nm$^{2}$ delimited by a closed line
of typically several tenths of atoms or molecules placed next to each other
one at a time on an atomically flat metallic surface using a STM. The same
microscope can be used to perform STS to study with meV resolution the
electronic density inside these corrals \cite{man,eig,cro,hel}. Particularly
interesting are the (111) surfaces of Cu, Au and Ag. These metals have
nearly spherical Fermi surfaces with eight necks at the [111] and equivalent
directions in which a gap opens. This allows the presence of Shockley
states localized at the (111) surface uncoupled to the bulk states for small
wave vector parallel to the surface and with nearly free electron dispersion 
\cite{hulb}. STS experiments performed on these surfaces reveal fascinating
standing-wave patterns and one can see the density coming from the wave
functions obtained solving the Schr\"{o}dinger equation for a
two-dimensional free electron gas inside a hard-wall corral \cite
{man,man2,cro,hel,fiete}. Experiments in which different atoms or molecules
were used to build the corral (Co, Fe, S, CO) suggest that the details of
the boundaries are not important for the physics. A continuous description
of the boundary is justified by the fact that the Fermi wave length $2\pi
/k_{F}\sim 3$ nm is larger than the distance between adatoms. However, as
discussed in section 6, the corrals are leaky and the hard-wall assumption
should be abandoned for a quantitative description.

The experiment of the quantum mirage is a beautiful combination of the
physics of the quantum corral and the many-body Kondo effect. One Co atom
acting as a magnetic impurity is placed at the focus of an elliptical corral
built on the Cu(111) surface, and a Fano dip is observed not only at the
place of the impurity, but remarkably also at the empty focus with reduced
magnitude \cite{man}. Variants of this experiment involving other corral
shapes and more that one impurity were presented in a conference \cite{man2}%
. In the original experiment, the space dependence of $\Delta dI/dV$, the
differential conductance after subtracting the corresponding result without
impurity, clearly resembles the density of the state number 42 in increasing
order of energy of free electrons in a hard-wall elliptical corral. This
suggests that the main features of this space dependence can be explained by
a one-body calculation. In fact, important features, like the possibility of
obtaining mirages out of the foci can be understood by a simple
tight-binding calculation \cite{wei} or from Green functions using hard-wall
eigenstates \cite{por}. Interesting effects like anti mirages were predicted
for a non-magnetic impurity inside a hard-wall elliptical corral \cite
{schmid}, and quantum mirages in s-wave superconductors were calculated \cite
{morr}. Also phenomenological scattering theories in which the energy
dependence of the Kondo resonance (directly related with the voltage
dependence of $dI/dV$) as well as an inelastic part of the scattering are
taken from experiment, are able to describe quantitatively the space
dependence \cite{fiete,agam,fie}. However, the calculation of the line shape
of $dI/dV$, its dependence on the particular geometry of the corral,
temperature or magnetic field, and the effects of interaction between
impurities is out of the scope of these one-body approaches.

The first calculation of the voltage dependence of $dI/dV$ has been done by
one of us using perturbation theory in the Coulomb repulsion of the Anderson
model $U$ \cite{rap,line} A many-body calculation of the mirage effect is a
challenge due to the particular nature of the conduction states brought by
the confinement in the corral. In particular, available exact results for
thermodynamic properties of the Kondo and Anderson impurities, obtained with
the Bethe ansatz, assume a constant density of conduction states \cite
{andrei,tsve,integ}, while the Wilson renormalization group \cite{wilson}
(which allows accurate calculation of dynamical properties and was used in
the context of nanoscopic systems and STS \cite{izum,costi,cor1,cor2,zhu}),
requires high symmetry around the impurity. If only a finite number of
hard-wall eigenstates with well defined energies are mixed with the impurity
(a problem that can be treated with exact diagonalization \cite{wil,hal}),
the line shape of $\Delta dI/dV$ becomes qualitatively wrong (see section 6).
The reason is that the separation of the relevant energy levels is large in comparison
with the Kondo temperature $T_{K}\sim 53$ K, while one knows that for a well
developed Kondo resonance to exist, $T_{K}$ should be larger than the
average separation of the relevant levels \cite{thi}. This points to the
need of including a finite width of the corral eigenstates, which become
resonances \cite{rap}. This need persists in presence of direct
hybridization of the impurity with bulk states as shown in section 6 \cite
{note3}. The width of the resonances cannot be too large because otherwise
the space dependence of the state 42 of the elliptical corral, observed in $%
dI/dV$ \cite{man} would be blurred.

A subject of current interest and debate in the literature is the relative
importance of the hybridization of the impurity with bulk $V_b$ and surface $%
V_s$ states. A first principles calculation seems not possible because of
the large supercells needed. They should contain more than 10 layers
perpendicular to the [111] direction in order for the Shockley surface
state to develop, and more than 100 atoms per layer to reach the dilute
limit of Co impurities on the surface \cite{andrea}. On the basis of the
rapid decay in $\Delta dI/dV$ as the STM tip is moved away from an impurity
on a clean (111) surface, 
and a jellium theory of Plihal and Gadzuk \cite{pli}, Knorr {\it et al.}
concluded that bulk states dominate the formation of the Kondo singlet \cite
{knorr,schn}. This is in agreement with tight-binding calculations \cite
{andrea}. However, recently Lin, Castro Neto and Jones, using a nearly free
electron approximation, including the effect of the gaps in the [111] and
equivalent directions and calculating the wave functions under an adequate
surface potential, concluded that the Kondo effect in the Cu(111) surface is
dominated by surface states \cite{lin}. They also obtained good agreement
with experiment for the distance dependence of the amplitude of $dI/dV$ and
its voltage dependence on top of the impurity. Using a similar approach, but
without attempting to solve the many-body problem, Merino and Gunnarsson
concluded that surface states play an important role in the differential
conductance for a system with a magnetic impurity on a clean (111) surface 
\cite{meri}. Therefore, the issue of the relative importance of $V_b$ and $%
V_s$ remains unclear. In contrast, in absence of the impurity, the relative
contribution of the surface states to the conductance (STM tip-substrate
hybridization) is known to be between 1/2 and 2/3 from experiments in which
the bias voltage is swept below the bottom of the surface band ($\sim 0.45$
eV below the Fermi energy) \cite{knorr,burg,jean}.

Since from the experiments we know that the presence of the corral strongly
affects electronic structure of the surface states, it is clear that the
variation of the line shape of $dI/dV$ for different corrals or positions of
the impurity inside the corral and its comparison with theory should help to
elucidate the relative role of the hybridization of the impurity with
surface and bulk states. A stronger sensitivity to the geometry implies a
greater participation of the surface states in the formation of the Kondo
resonance. Also the interaction between magnetic impurities inside a quantum
corral should increase with the relative importance of surface states \cite
{wil}. Unfortunately only the voltage dependence of $dI/dV$ for a Co atom on
a clean Cu(111) surface and on an elliptical corral built on that surface is
available for comparison \cite{man}. Using perturbation theory in $U$ both
line shapes are qualitatively explained without bulk states \cite{line} (see
section 8). However, as we will show, this seems to be a particular case and
usually the shape and width of the Fano dip are more sensitive to the
geometry.

In this work we discuss the main aspects of the physics of the quantum
mirage. The emphasis is on the basic understanding of the phenomenon and its
many-body aspects rather than on quantitative fits. The latter would require
more detailed knowledge of matrix elements and their wave vector dependence,
crystal fields and other details. We extend previous many-body
calculations for the space and voltage dependence of $\Delta dI/dV$ to new
different situations. This can serve as a basis for comparison with
experiment and help to elucidate the relative participation of surface and bulk
states in the formation of the Kondo singlet for a Co atom on a Cu(111)
surface. We use three different many-body techniques: perturbation theory
in $U$ \cite{yos,hor}, exact diagonalization plus embedding \cite
{fer,buss,wil2} and a slave-boson mean-field approximation (SBMFA) \cite
{colb,hewson,kang,lady,ihm}. The former two have already been applied to the
quantum mirages \cite{rap,line,lob1,lob2,wil} but have the disadvantage that
they do not reproduce the correct exponential dependence of $T_{K}$ with the
coupling constant for large $U/\Delta $, where $\Delta $ is the resonance
level width \cite{note}. Therefore, the SBMFA is more appropriate to study
the dependence of the width of the resonance on geometry.

The paper is organized as follows. In section 2 we present the impurity
Anderson model for either the corral or open surfaces, and discuss its
assumptions and limitations. Section 3 discusses the Kondo resonance and
Fano antiresonance in the simplest version of the model for later
comparison. The formalism and basic equations that determine the tunneling
current are presented in section 4, using a many-body formalism, including
tunneling of the tip of the STM with surface, bulk and impurity states.
Section 5 is rather technical and explains the different many-body
approaches. In section 6 we explain the effects of the confinement on the
surface states, and how they are transmitted to the Kondo resonance and the
line shape of the mirage effect. Section 7 is devoted to the space
dependence of the differential conductance $dI/dV$ inside an elliptical
quantum corral, the effect of the impurity on it ($\Delta dI/dV$), and the
relation of these quantities with the wave functions of the surface states
inside the corral. This brings insight into the effect of the width of the
surface states and what controls the intensity at the mirage point. In
section 8 we present results for the dependence of $\Delta dI/dV$ on bias
voltage in different situations: clean surface, elliptical corrals and a
circular corral. In section 9 we estimate a lower bound for the
participation of surface states in the Kondo resonance. In section 10 the
interaction between two Anderson impurities inside an elliptical corral is
discussed. Section 11 contains a summary and a discussion.

\section{The model}

In this section, we explain and discuss the model used to describe the
electronic structure of a system composed of one magnetic impurity
interacting with surface and bulk states. The case of two impurities is left
for section 10. The surface states can correspond to eigenstates of a clean
perfect surface, or to a surface with a soft-wall corral. In both cases, the
energy spectrum of the surface states is continuous. The wave functions $%
\varphi _{j}(r)$ of the surface eigenstates are normalized in a large area 
\cite{lob1,lin}. Of course, all physical results are independent of this
area.

We take only one localized $d$ orbital for the impurity. Technically it
renders some many-body techniques easier (except the SBMFA). Previously
\'{U}js\'{a}ghy {\it et al.} \cite{uj} assumed a fully degenerate ground
state while other recent calculations for impurities on (111) surfaces
considered the $d_{3z^{2}-r^{2}}$ orbital more important \cite
{lin,meri,meri2}. Tight binding calculations suggest that Cu(111) surface
states hybridize more strongly with the Co $3d_{3z^{2}-r^{2}}$ orbital,
while bulk states prefer $3d_{xz}$ and $3d_{yz}$ orbitals \cite{andrea}.
Recent accurate calculations using the Wilson renormalization group indicate
that for the expected filling of near one $d$ hole per Co impurity, the
Kondo resonance becomes strongly asymmetric in the orbitally degenerate case 
\cite{zhu}. Then, one would expect in this case also a strongly asymmetric 
$\Delta dI/dV$ in contrast to the experimental observations for 
the (111) surface \cite
{man,knorr}. We neglect the $s,p$ orbitals of the impurity.

The Hamiltonian can be written as 
\begin{eqnarray}
H &=&\sum_{j\sigma }\varepsilon _{j}^{s}s_{j\sigma }^{\dagger }s_{j\sigma
}+\sum_{j\sigma }\varepsilon _{j}^{b}b_{j\sigma }^{\dagger }b_{j\sigma
}+E_{d}\sum_{\sigma }d_{\sigma }^{\dagger }d_{\sigma }+Ud_{\uparrow
}^{\dagger }d_{\uparrow }d_{\downarrow }^{\dagger }d_{\downarrow }  \nonumber
\\
&&+\sum_{j\sigma }(V_{s}^{j}d_{\sigma }^{\dagger }s_{j\sigma }+\text{H.c.}%
)+\sum_{j\sigma }(V_{b}^{j}d_{\sigma }^{\dagger }b_{j\sigma }+\text{H.c.}).
\label{ham}
\end{eqnarray}
where $s_{j\sigma }^{\dagger }$ ($b_{j\sigma }^{\dagger }$) are creation
operators for an electron in the $j^{th}$ surface (bulk) conduction
eigenstate in the absence of the impurity, but including the corral if
present. The impurity is placed at the two-dimensional position $R_{i}$ on
the surface, and we assume that the hybridization of the impurity $d$
orbital with the surface state $j$ is proportional to its normalized wave
function at that point $\varphi _{j}(R_{i})$ \cite{por,rap}. Similarly for
the bulk states, the hybridization is proportional to some average $\psi
_{j}(R_{i})$ of the bulk wave function in the  direction normal to the
surface, that depends on $R_{i}$:

\begin{equation}
V_{s}^{j}=V_{s}\lambda \varphi _{j}(R_{i}),\text{ }V_{b}^{j}=V_{b}\psi
_{j}(R_{i}),  \label{vsb}
\end{equation}
where $V_{s}$, $V_{b}$ are energies representing local hybridizations in a 
tight-binding model \cite{wei,line} (see next section) and $\lambda =2.38$ \AA\ is the square root
of the surface per Cu atom of a Cu(111) surface. We assume also a constant
density of bulk states. However, we must warn that recent calculations
obtain a significant dependence of the matrix elements with wave vector \cite
{lin,meri,meri2}. This dependence affects the line shape of $dI/dV$.
Nevertheless, one expects that the {\em trends} in the modifications of the
voltage dependence of $dI/dV$ due to the modifications of the geometry
remain the same, at least on a qualitative level. This is also suggested by
the weak dependence of the results on the cutoff for the surface states $%
E_{c}$, which should be introduced in any theory of quantum corrals to avoid
divergences in the Green's functions for the surface states. This can be
though as an energy dependent hybridization $V_{s}$ which is constant below $%
E_{c}$ and goes to zero abruptly at $E_{c}$. A linearly decreasing $V_{s}$
has also been used \cite{rap}. Increasing $E_{c}$ leads to a weak increase
in the width of the resonances and to a more asymmetric line shape, but the
main conclusions regarding the mirage effect are not altered.

The many-body part of the Hamiltonian which renders it non-trivial in all
cases is the on-site Coulomb repulsion at the impurity site $U$. Another
difficulty is the calculation of the surface wave functions $\varphi _{j}(r)$
for soft walls. They have been calculated exactly for a soft circular corral 
\cite{lob1,lob2} and can be reasonably well approximated for an elliptical
corral \cite{lob1}. We return to this point in section 6. For open
structures, the surface Green's function can be calculated using scattering
theory \cite{fiete}. However, this renders the many-body problem too
difficult.

\section{Simple picture of the Kondo resonance and Fano antiresonance}

Before discussing the many-body techniques and the effect of the corral, we
want to illustrate some basic features of the Kondo physics, using the
simplest case of the Anderson model in which the impurity is hybridized with
only one band (either surface or bulk) with constant density of states $\rho
_{0}$ and wave vector independent hybridization. The Hamiltonian is given by
Eq.  (\ref{ham}) eliminating the terms with $s_{j\sigma }$, 
considering that
the $b_{j\sigma }^{\dagger }$ create Bloch waves, ($b_{{\boldmath k}\sigma
}^{\dagger }
=(1/\sqrt{N})\sum_{l}\exp (-i{\boldmath k}\cdot {\boldmath R}_{l})b_{l\sigma
}^{\dagger }$, where $b_{l\sigma }^{\dagger }$ creates an electron at site $l$ 
with position ${\boldmath R}_{l}$) and taking $V_{b}^{j}=V/\sqrt{N}$ where $N$
is the number of sites. Then, the impurity is hybridized with the band at
one site that we call $i$.

The impurity spectral density $\rho _{d\sigma }(\omega )$ of this model has
been calculated accurately using Wilson renormalization group \cite{costi3}
and agree qualitatively with those of perturbation theory in $U$ \cite{hor}.
The results presented in Fig. 1 (a) were obtained using a self-consistent
approach \cite{levy} based on an interpolation for the self energy of the
Green's function between the expression up to second order perturbation
theory in the Coulomb repulsion $U$ \cite{yos,hor} and the exact result for $%
U\rightarrow \infty $. The resonant level width is $\Delta =\pi \rho
_{0}V^{2}$. This approximation works well for $U\leq 8\Delta $ \cite{pro}.

\begin{figure}[h!]
\hskip2.0cm\psfig{file=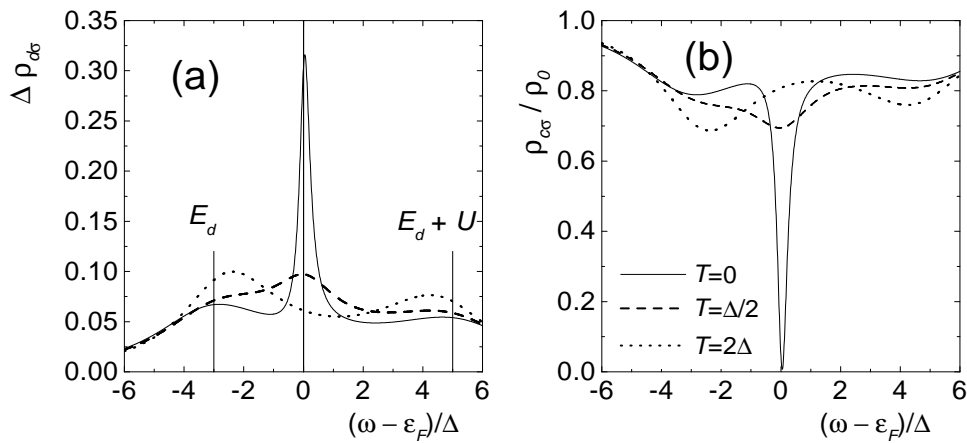,width=14.0cm,silent=} 
\vskip0.4cm
\caption{(a) spectral density of an impurity hybridized with a
featureless conduction band and (b) spectral density at the conduction site
hybridized with the impurity, as a function of energy at different
temperatures. Parameters are $E_{d}-\epsilon _{F}=-3\Delta $ and $U=8\Delta $.}
\end{figure}

As seen in Fig. 1 (a), $\rho _{d\sigma }(\omega )$ shows characteristic
charge fluctuation peaks (or shoulders for small $U$) at $E_{d}$ and $%
E_{d}+U $ and another peak near the Fermi energy $\epsilon _{F}$
characteristic of the Kondo regime. This peak is the so called Kondo
resonance. Its half width at half maximum corresponds to the Kondo
temperature $T_{K}$. At temperatures above $T_{K}$ the Kondo effect
disappears and the spectral weight of the Kondo peak is transferred to the
other two.

The STS is much more sensitive to the conduction electrons than to the
localized ones because the former are more extended in space and reach the
tip of the STM with a larger amplitude. The site most affected by the
impurity is the one which hybridizes with it ($i$). Using equations of
motion (in the same way as done in the next section), it is easy to show that the
Green function at this site is (the spin index is dropped for simplicity)

\begin{equation}
G_{ii}(\omega )=G_{ii}^{0}(\omega )+\left[ VG_{ii}^{0}(\omega )\right]
^{2}G_{d}(\omega ),  \label{gsimple}
\end{equation}
where $G_{ii}^{0}(\omega )$ is the corresponding Green's function in the
absence of the impurity and $G_{d}(\omega )$ is the Green's function of the
impurity. This equation is exact and does not depend on the approximations
for $G_{dd}(\omega )$. If the unperturbed conduction band extends from $-W$
to $W$ then

\begin{equation}
G_{ii}^{0}(\omega )=\sum_{k}\frac{1}{\omega +i\eta -\epsilon _{k}}=\rho
_{0}\left[ \ln |\frac{W+\omega }{W-\omega }|-i\pi \Theta (W-|\omega
|)\right] .  \label{g0simple}
\end{equation}
If (as usual) $W\gg $ $T_{K}$ and we are interested in energies $|\omega
-\epsilon _{F}|\sim T_{K}$, we can neglect the first term inside the
brackets and approximate $G_{ii}^{0}(\omega )\cong -i\pi \rho _{0}$.
Replacing this in Eq. (\ref{gsimple}) and using $\rho _{i\sigma }(\omega )=-%
\mathop{\rm Im}%
[G_{ii}(\omega )]/\pi $ one obtains the very simple result

\begin{equation}
\frac{\rho _{i\sigma }(\omega )}{\rho _{0}}=1-\pi \rho _{d\sigma }(\omega
)\Delta .  \label{fano}
\end{equation}
Thus a peak in $\rho _{d\sigma }(\omega )$ implies a dip in the conduction
density of states and more pronounced near the impurity (see Fig. 1 (b)). In
more complex situations, in particular when confinement due to the corral is
important, the real part of $G_{ii}^{0}(\omega )$ cannot be neglected and
the dip in $\rho _{i\sigma }(\omega )$ is not directly related with the
Kondo peak in $\rho _{d\sigma }(\omega )$. In extreme cases, either the dip
is replaced by a peak \cite{lin} or the structure near $\epsilon _{F}$
disappears, as we will show in section 6.

In any case, the above simple picture corresponds to a first rough
approximation of the experimental observations of the voltage dependence of $%
dI/dV$ for impurities on the (111) surfaces of Cu and noble metals, and we
will use it for later comparison.

\section{The tunneling conductance}

In this section we write the basic equations which relate $dI/dV$ with the
Green's function at the impurity site. We include the hopping of the tip of
the STM with the impurity, surface and bulk states in a many-body formalism.
The tunneling geometry and energy diagram is shown for example in Fig. 1 of
Ref. \cite{fiete}, but the impurity should be included if the tunneling
current is measured near it \cite{schi}, and also the bulk states according
to experiment \cite{knorr,burg,jean}.

The total system $S$ consists of a subsystem $S_{H}$ described by the
Hamiltonian $H$ [Eq. (\ref{ham})] and $S_{t}$ contains the tip, which we
assume can be described as a non-interacting system with one-particle
energies $\epsilon _{k}$ and the Fermi energy set at zero. $S_{H}$ has all
one particle energies, including the Fermi level, displaced by $eV$ to lower
energies by a bias voltage $V$, where $e$ is the elementary charge. For
simplicity we treat the case of positive $V$ in which electrons are
transferred from the tip to $S_{H}$. Extension to negative $V$ is trivial
using an electron-hole transformation. We assume a local hopping of the tip
with the different states

\begin{equation}
H_{mix}=A\sum_{k\sigma }(t_{k\sigma }^{\dagger }h_{\sigma }+\text{H.c.}),
\label{hmix}
\end{equation}

\begin{equation}
h_{\sigma }=\lambda \sum_{j}\varphi _{j}(R_{t})s_{j\sigma }+p\sum_{j}\psi
_{j}(R_{t})b_{j\sigma }+q(|R_{t}-R_{i}|)d_{\sigma }.  \label{h}
\end{equation}
Here $t_{k\sigma }^{\dagger }$ creates an electron in the tip eigenstate $k$
with spin $\sigma $, $R_{t}$ describes the coordinates of the tip on the
plane, and $A$, $p$, and $q$ are parameters that describe the hopping of the
tip with the different states of $S_{H}$. The function $q(|R_{t}-R_{i}|)$ is
small and decays strongly with the distance between the tip and the impurity 
$|R_{t}-R_{i}|$ due to the strongly localized nature of the impurity wave
function. However, when $R_{t}-R_{i}=0$, a small $q$ introduces an important
source of asymmetry in the line shape of $dI/dV$ in addition to that
corresponding to the structure of the Green's functions (see section 8).

Treating $H_{mix}$ in lowest order in perturbation theory and at $T=0$,
using Fermi's golden rule, the current due to the transfer of electrons from 
$S_{t}$ to $S_{H}$ becomes

\begin{equation}
I=\frac{2\pi e}{\hbar }A^{2}\sum_{\nu }|\langle \nu |\sum_{k\sigma
}h_{\sigma }^{\dagger }t_{k\sigma }|g\rangle |^{2}\delta (E_{\nu }-E_{g}-eV),
\label{i}
\end{equation}
where $|\nu \rangle $, $E_{\nu }$ are the eigenstates and energies of $S$
and $|g\rangle $ is the ground state assumed non degenerate. Using the same
notation with a subscript $H$ for $S_{H}$, we have $|g\rangle
=\prod_{k\sigma }t_{k\sigma }^{\dagger }|g_{H}\rangle $ with the product
restricted to $k$ such that $\epsilon _{k}<\epsilon _{F}=0$. Replacing above
and doing the calculations within $S_{t}$ one has

\begin{equation}
I=\frac{2\pi e}{\hbar }A^{2}\sum_{\nu _{H}}\sum_{k\sigma }^{\prime }|\langle
\nu _{H}|h_{\sigma }^{\dagger }|g_{H}\rangle |^{2}\delta (E_{\nu
_{H}}-\epsilon _{k}-E_{g_{H}}-eV).  \label{i2}
\end{equation}
Using the Lehman representation \cite{mahan}, the sum over $\nu _{H}$ is
seen to represent the part of the spectral density $\rho _{h\sigma
}(\epsilon _{k}+eV)$ of $h_{\sigma }^{\dagger }$ for electron addition. This
corresponds to excitations above $\epsilon _{F}$ with positive argument of $%
\rho _{h\sigma }$ \cite{wagner}. By symmetry it is independent of $\sigma $.
Transforming the sum over the tip states $k$ as an integral assuming a
constant density of states $\rho _{t}$ one gets

\begin{equation}
I\left( \frac{4\pi e}{\hbar }A^{2}\rho _{t}\right) ^{-1}=\int_{-eV}^{0}\rho
_{h\sigma }(\epsilon +eV)d\epsilon =\int_{0}^{eV}\rho _{h\sigma }(\omega
)d\omega \text{.}  \label{i3}
\end{equation}
From here, it is clear that the differential conductance is proportional to
the spectral density of the state $h_{\sigma }(R_{t})$:

\begin{equation}
dI/dV\sim \rho _{h\sigma }(eV)=-\frac{1}{\pi }%
\mathop{\rm Im}%
G_{h\sigma }(eV),  \label{didv}
\end{equation}
where $G_{h\sigma }(\omega )=\langle \langle h_{\sigma };h_{\sigma
}^{\dagger }\rangle \rangle _{\omega }$ is the Green's function of $%
h_{\sigma }(R_{t})$. Therefore, in the rest of the paper we will be mainly
concerned on the space and energy dependence of $\rho _{h\sigma }$. This
spectral density can be related with the Green's function for the $d$
electrons $G_{d\sigma }(\omega )=\langle \langle d_{\sigma };d_{\sigma
}^{\dagger }\rangle \rangle _{\omega }$, and the unperturbed Green's
functions for conduction electrons using equations of motion. Writing $%
c_{j\sigma }$ to represent either $s_{j\sigma }$ or $b_{j\sigma }$, the
relevant equations can be written in the form

\begin{eqnarray}
(\omega -\epsilon _{j}^{c^{\prime }})\langle \langle c_{j\sigma }^{\prime
};c_{j\prime \sigma }^{\dagger }\rangle \rangle _{\omega } &=&\delta
_{jj\prime }\delta _{cc\prime }+\bar{V}_{c^{\prime }}^{j}\langle \langle
d_{\sigma };c_{j\prime \sigma }^{\dagger }\rangle \rangle _{\omega }, 
\nonumber \\
(\omega -\epsilon _{j}^{c})\langle \langle c_{j\sigma };d_{\sigma }^{\dagger
}\rangle \rangle _{\omega } &=&\bar{V}_{c}^{j}\langle \langle d_{\sigma
};d_{\sigma }^{\dagger }\rangle \rangle _{\omega },  \nonumber \\
(\omega -\epsilon _{j}^{c})\langle \langle d_{\sigma };c_{j\sigma }^{\dagger
}\rangle \rangle _{\omega } &=&V_{c}^{j}\langle \langle d_{\sigma
};d_{\sigma }^{\dagger }\rangle \rangle _{\omega }.  \label{m}
\end{eqnarray}
Dropping the spin indices, using these equations and introducing the
non-interacting Green's functions (in absence of the impurity) for
conduction electrons

\begin{eqnarray}
G_{s}^{0}(R_{1},R_{2},\omega ) &=&\langle \langle \sum_{j}\varphi
_{j}(R_{1})s_{j\sigma };\overline{\varphi }_{j^{\prime }}(R_{2})s_{j\prime
\sigma }^{\dagger }\rangle \rangle _{\omega }=\sum_{j}\frac{\varphi
_{j}(R_{1})\overline{\varphi }_{j}(R_{2})}{\omega -\epsilon _{j}^{s}}, 
\nonumber \\
G_{b}^{0}(R_{1},R_{2},\omega ) &=&\sum_{j}\frac{\psi _{j}(R_{1})\overline{%
\psi }_{j}(R_{2})}{\omega -\epsilon _{j}^{b}},  \label{g0}
\end{eqnarray}
the Green's function for the $h$ operators becomes

\begin{eqnarray}
G_{h}(R_{t},R_{i},\omega ) &=&\lambda ^{2}G_{s}^{0}(R_{t},R_{t},\omega
)+p^{2}G_{b}^{0}(R_{t},R_{t},\omega )+\Delta G_{h}(R_{t},R_{i},\omega ), 
\nonumber \\
\Delta G_{h}(R_{t},R_{i},\omega ) &=&F(R_{t},R_{i},\omega
)F(R_{i},R_{t},\omega )G_{d}(\omega ),  \nonumber \\
F(R_{1},R_{2},\omega ) &=&V_{s}\lambda ^{2}G_{s}^{0}(R_{1},R_{2},\omega
)+pV_{b}G_{b}^{0}(R_{1},R_{2},\omega )+q(|R_{1}-R_{2}|).  \label{gh}
\end{eqnarray}
Here, the first two terms when replaced in Eq. (\ref{didv}) describe $dI/dV$
in the absence of the impurity, while $\Delta G_{h}$ describes the effect of
the impurity on the differential conductance $\Delta dI/dV$.

Note that the space dependence of $dI/dV$ {\em is determined only by the
non-interacting conduction electron Green's functions}. In particular at a
distance of the impurity larger than $\sim $0.5 nm, $q(|R_{t}-R_{i}|)$
becomes irrelevant, the bulk part becomes less important in $\Delta dI/dV$
due to its more rapid decay with the distance between the tip and the
impurity $|R_{t}-R_{i}|$ \cite{note5}, and the space dependence is dominated
by $G_{s}^{0}(R_{t},R_{i},\omega )$. The impurity Green's function $G_{d}$
can only alter the relative weight of the real and imaginary part of the
other factors in $\Delta G_{h}$. There is a natural length scale in the
Kondo problem $\xi =\hbar v_{F}/T_{K}$, where $v_{F}$ is the Fermi velocity.
It has been interpreted as the size of the cloud of conduction electrons
that screen the localized spin in the Kondo effect. The existence of this
cloud is still controversial \cite{sor,barz,col}. Theoretical work has shown
that the persistent current as a function of flux $j(\Phi )$ in mesoscopic
rings with quantum dots changes its shape smoothly as the length of the ring 
$L$ goes through $\xi $ and that $jL$ is a universal function of $L/\xi $ 
\cite{aff,pc}. However, in our case, it is clear that $\xi $ plays no role
in the space dependence of $dI/dV$.

\section{The many-body techniques}

The core of the many-body problem is to solve the impurity Green's function $%
G_{d\sigma }$ which enters Eq. (\ref{gh}) and determines $dI/dV$ through Eq.
(\ref{didv}). Here we present results using three different techniques: a)
perturbation theory up to second order in $U$, b) slave-boson mean-field
approximation (SBMFA) and c) exact diagonalization plus embedding (EDE). The
first one has been already used by us to study the mirage effect \cite
{rap,line,lob1,lob2} and by one of us \cite{line} and Shimada {\it et al.} 
\cite{shima} to study the line shape of $dI/dV$ in absence of the corral.
The latter problem was also studied recently using the SBMFA \cite{lin}, and
to the best of our knowledge the results presented in section 8 are the
first application of this technique to the mirage effect. EDE has been used
in Ref. \cite{wil}.

\subsection{Perturbation theory in the Coulomb repulsion}

The starting point is the calculation of the non-interacting problem ($U=0$)
but with $E_{d}$ replaced by effective one-particle $d$ level $E_{d\sigma
}^{eff}$. Using equations of motion similar to Eqs. (\ref{m}), and assuming $%
V_{b}^{2}G_{b}^{0}(R_{i},R_{i},\omega )=-i\delta _{b}$ independent of $%
\omega $ (as in the simple case of section 3), the resulting non-interacting
impurity Green's function becomes:

\begin{equation}
G_{d\sigma }^{0}(\omega )=\frac{1}{\omega -E_{d\sigma }^{eff}+i\delta
_{b}-(V_{s}\lambda )^{2}G_{s}^{0}(R_{i},R_{i},\omega )}.  \label{god}
\end{equation}
The first choice for $E_{d\sigma }^{eff}$ would be the Hartree-Fock value 
$E_{d\sigma 0}^{eff}=E_{d}+U\langle d_{\bar{\sigma}}^{\dagger }d_{\bar{\sigma}}
\rangle $ \cite{hor}. However, out of the symmetric case $%
E_{d}+U/2=\epsilon _{F}$, better results are obtained if $E_{d\sigma }^{eff}$
and $\langle d_{\sigma }^{\dagger }d_{\sigma }\rangle $ are calculated
self-consistently using interpolative schemes that reproduce correctly the
physics not only for small $U$ but also for infinite $U$ \cite
{pro,levy,pc,kaj}. For example, the persistent current in small rings with
quantum dots practically coincides with exact results for $U\sim 6\Delta $,
where $\Delta $ is the resonant level width \cite{pc}. At the symmetric case
the theory is quantitatively correct up to $U\sim 8\Delta $ \cite{sil}. To
avoid self-consistency we take parameters near the symmetric case, for which $%
E_{d\sigma }^{eff}$ is near the Fermi energy $\epsilon _{F}$. This is
consistent with first-principle calculations \cite{llois}.

\begin{figure}[h!]
\hskip2.0cm\psfig{file=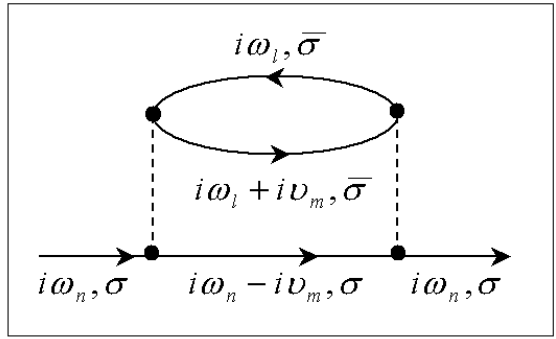,width=10.0cm,silent=} 
\vskip0.4cm
\caption{Feynman diagram for the contribution to the self energy of the 
$d$ electrons in second order in the Coulomb repulsion $U$.}
\end{figure}

The interacting impurity Green's function can be written in the form

\begin{equation}
G_{d\sigma }^{-1}(\omega )=\left[ G_{d\sigma }^{0}(\omega )\right]
^{-1} + E_{d\sigma}^{eff}-E_{d\sigma 0}^{eff}-\Sigma _{\sigma }(\omega ),  \label{g}
\end{equation}
and the approximation consists in calculating $\Sigma _{\sigma }(\omega )$
in second-order perturbation theory in $U$ \cite{yos,hor} (the first-order
terms are already included in $E_{d\sigma 0}^{eff}$). The corresponding
Feynman diagram is shown in Fig. 2. Using the analytical extension of the
time ordered $G_{d\sigma }^{0}(\omega )$ to Matsubara frequencies, the
expression for the self energy reads 
\begin{eqnarray}
\Sigma _{\sigma }(i\omega _{n},T) &=&U^{2}T\sum_{m}G_{d\sigma }^{0}(i\omega
_{n}-i\nu _{m})\chi (i\nu _{m})\text{;}  \nonumber \\
\chi (i\nu _{m}) &=&-T\sum_{l}G_{d\bar{\sigma}}^{0}(i\omega _{l})G_{d\bar{%
\sigma}}^{0}(i\omega _{l}+i\nu _{m}),  \label{sigma}
\end{eqnarray}
where the $\omega _{i}$ ($\nu _{m}$) are fermionic (bosonic) frequencies.
The evaluation of the Matsubara sums is greatly facilitated by the fact that
the unperturbed Green's function for surface states $G_{s}^{0}(R_{1},R_{2},%
\omega )$, which for a soft corral involve a continuous distribution of
energy, can be well approximated by a sum over a finite number of simple
fractions with simple poles in the complex plane [see Eq. (\ref{goap}) of
section 6] \cite{lob1}.

\vskip1.0cm
\begin{figure}[h!]
\hskip2.0cm\psfig{file=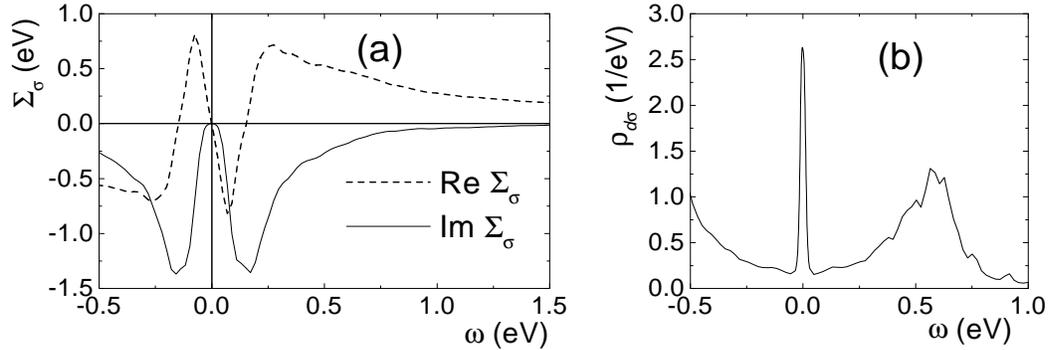,width=14.0cm,silent=} 
\vskip0.4cm
\caption{(a) Self energy and (b) impurity spectral density as a function
of energy for a system with an impurity placed at position $(-0.4a,0)$
inside an elliptical quantum corral with eccentricity 1/2 and size such that
the state 35 is at the Fermi level. Parameters are in the text (section 5 A).}
\end{figure}

In Fig. 3 (a) we show the resulting $\Sigma _{\sigma }(\omega )$ at zero
temperature for an elliptical corral with eccentricity $e=1/2$, like that
used in the experiment of Manoharan {\it et al.} \cite{man}, but with
semimajor axis reduced to $a=6.46$ nm so that the state 35 of the hard-wall
corral falls at the Fermi level. The impurity was placed at a maximum of the
wave function of this state ($x=\pm 0.4a$, $y=0$, see Fig. 10). For
simplicity we took $V_{b}=0$, and $G_{s}^{0}(R_{1},R_{2},\omega )$ was
constructed from hard-wall eigenstates broadened by an imaginary part 
$\delta =20$ meV [Eq. (\ref{goap}) ]. For the hybridization with surface
states we took an energy dependent decreasing function $V_{s}(\epsilon
)=0.67 $eV$\max \left( 1+\frac{\epsilon _{F}-\epsilon }{\text{eV}},0\right) $
\cite{rap}, which leads to a more symmetric impurity density of states $\rho
_{d\sigma }(\omega )$. The zero of energy is set at $\epsilon _{F}=0$, and $%
E_{d\sigma }^{eff}=22$ meV. We took $U=1$ eV. While $U=2.84$ eV has been
estimated \cite{uj}, this approximation ceases to be reliable for larger
values of $U$ \cite{note6}. The imaginary part of $\Sigma _{\sigma }(\omega )$ vanishes
at $\epsilon _{F}$ and has a quadratic dependence on energy near $\epsilon
_{F}$ , respecting Fermi liquid properties \cite{lang}.

The particular structure of $\Sigma _{\sigma }(\omega )$ near $\epsilon _{F}$
leads to the development of the Kondo peak in the impurity spectral density $%
\rho _{d\sigma }(\omega )=-\frac{1}{\pi }
{\rm Im} G_{d\sigma }(\omega )$. 
This function is shown in Fig. 3 (b) for a range of
energies extending between the bottom of the surface band and the smooth
cutoff in the hybridization. The overall structure is similar to that shown
in Fig. 1 (a), with two charge fluctuation peaks and the Kondo peak.
However, the uneven structure of the conduction band, which in this case
is a sum of broadened peaks rather than a flat band, introduces some
wiggles. This is particularly clear for the charge fluctuation peak near $%
E_{d}+U$ $\sim 0.5$ eV. The effects of the confinement will be discussed in
the next section.

Unless otherwise indicated, the results presented in this paper were
obtained by this technique.

\subsection{The slave-boson mean field approximation (SBMFA)}

This approximation for the $U\rightarrow \infty $ limit of the Anderson
model is in some sense a complement of the previous one, which is valid for
small or moderate $U$ \cite{note6}. The slave-boson representation consists in writing $%
d_{\sigma }^{\dagger }=f_{\sigma }^{\dagger }b$ as a product of a fermion
operator $f_{\sigma }^{\dagger }$ and a bosonic one $b$ \cite{colb}. For $%
U\rightarrow \infty $, double occupancy is forbidden and this is expressed
by the constraint $b^{\dagger }b+\sum_{\sigma }f_{\sigma }^{\dagger
}f_{\sigma }=1$ introduced by a Lagrange multiplier $\Lambda $ in the
Hamiltonian $H$ \cite{colb,newns}. We present the formalism for the 
$SU(N)$ generalization of our model, in which the index $\sigma $ can run
over a set of $N$ degenerate states (instead of only 2). In mean field, the
bosonic operators are replaced by a number $b\rightarrow \langle b\rangle
=\langle b^{\dagger }\rangle =r$, and $r$ and $\Lambda $ are obtained
minimizing the free energy of the resulting model for free fermions. In this
approximation, the charge fluctuation peaks (at $E_{d}$ and $E_{d}+U$) are
absent in the spectral density. However, in the
Kondo regime, for zero or small temperature and energies near the Fermi
energy, the approximation seems to be reliable \cite{hewson}. We restrict
our calculations to $T=0$.

In the SBMFA, the impurity Green's function near the Fermi energy $\epsilon
_{F}=0$ is just

\begin{equation}
G_{d\sigma }(\omega )=r^{2}G_{f\sigma }(\omega ),  \label{gdf}
\end{equation}
and the Green's function $G_{f\sigma }(\omega )=\langle \langle f_{\sigma
};f_{\sigma }^{\dagger }\rangle \rangle _{\omega }$ is obtained solving the
following effective Hamiltonian, which results from Eq. (\ref{ham}) with the
above explained replacements

\begin{eqnarray}
H_{eff} &=&\sum_{j\sigma }\varepsilon _{j}^{s}s_{j\sigma }^{\dagger
}s_{j\sigma }+\sum_{j\sigma }\varepsilon _{j}^{b}b_{j\sigma }^{\dagger
}b_{j\sigma }+(E_{d}+\Lambda )\sum_{\sigma }f_{\sigma }^{\dagger }f_{\sigma
}+\Lambda (r^{2}-1)  \nonumber \\
&&+r\left[ \sum_{j\sigma }(V_{s}^{j}f_{\sigma }^{\dagger }s_{j\sigma }+\text{%
H.c.})+\sum_{j\sigma }(V_{b}^{j}f_{\sigma }^{\dagger }b_{j\sigma }+\text{H.c.%
})\right] .  \label{heff}
\end{eqnarray}
Minimization of the energy $\langle H_{eff}\rangle $ with respect to $%
\Lambda $ leads to

\begin{equation}
r^{2}=1-\sum_{\sigma }n_{\sigma }=1-Nn_{\sigma },  \label{r2}
\end{equation}
where in the second equality we assume $SU(N)$ invariance and

\begin{equation}
n_{\sigma }=\langle f_{\sigma }^{\dagger }f_{\sigma }\rangle =-\frac{1}{\pi }%
\int d\omega f(\omega ) {\rm Im} G_{f\sigma }(\omega ),  \label{ns}
\end{equation}
where $f(\omega )$ is the Fermi function.

Using the Hellmann-Feynman theorem \cite{hell}, the other equation to be
solved self-consistently reads

\begin{equation}
\frac{1}{2r}\frac{\partial \langle H_{eff}\rangle }{\partial r}=\Lambda
+S+B=0,  \label{dr}
\end{equation}

\begin{equation}
S=\frac{1}{2r}\sum_{j\sigma }\langle V_{s}^{j}f_{\sigma }^{\dagger
}s_{j\sigma }+\text{H.c.}\rangle ,\text{ }B=\frac{1}{2r}\sum_{j\sigma
}\langle V_{b}^{j}f_{\sigma }^{\dagger }b_{j\sigma }+\text{H.c.}\rangle .
\label{esb}
\end{equation}
The expectation values entering this equation can be evaluated as integrals
over $f(\omega )$ times the imaginary part of Green's functions of the same
form of the first member of the second and third of the Eqs. (\ref{m}) \cite
{zuba}. From the differences between $H$ (Eq. (\ref{ham})) and $H_{eff}$
(Eq. (\ref{heff})) one sees that these equations can be used  with $d_{\sigma
}$ replaced by $f_{\sigma }$ and a factor $r$ multiplying $V_{s}^{j}$ and $%
V_{b}^{j}$. Then

\begin{equation}
S=-\frac{1}{\pi }
{\rm Im}
\left[ \int d\omega f(\omega ) 
\sum_{j\sigma }\frac{|V_{s}^{j}|^{2}}{\omega +i\eta -\epsilon
_{j}^{s}}G_{f\sigma }(\omega )\right] =-\frac{N}{\pi } 
V_{s}^{2}\lambda ^{2}
{\rm Im}
\left[ \int d\omega f(\omega )G_{s}^{0}(R_{i},R_{i},\omega )G_{f\sigma
}(\omega )\right] ,  \label{ss}
\end{equation}
where for the last equality we used Eqs. (\ref{vsb}) and (\ref{g0}). In a
similar way one has

\begin{equation}
B=-\frac{N}{\pi }V_{b}^{2}%
\mathop{\rm Im}%
\left[ \int d\omega f(\omega )G_{b}^{0}(R_{i},R_{i},\omega )G_{f\sigma
}(\omega )\right] ,  \label{bb}
\end{equation}
while the $f$ electron Green's function is:

\begin{equation}
G_{f\sigma }(\omega )=\frac{1}{\omega -E_{d}-\Lambda -(rV_{s}\lambda
)^{2}G_{s}^{0}(R_{i},R_{i},\omega )-(rV_{b})^{2}G_{b}^{0}(R_{i},R_{i},\omega
)}.  \label{gf}
\end{equation}

From the self-consistent solution of Eqs. (\ref{r2}) to (\ref{gf}) we obtain 
$G_{f\sigma }(\omega )$ and $r$. The differential conductance is then
obtained using Eqs. (\ref{didv}), (\ref{gh}) and (\ref{gdf}). For the
calculations shown here, we take $N=2$ because otherwise the line shape
becomes too asymmetric in comparison with experiments for the expected 3d$%
^{9}$ configuration of the Co impurity \cite{zhu}.

In the absence of the corral, for an impurity on a clean surface, we assume
constant symmetric density of states as in Eq. (\ref{g0simple}):

\begin{equation}
\lambda ^{2}G_{s}^{0}(R_{i},R_{i},\omega )=\rho _{s}\left[ \ln \left( \frac{%
W_{s}+\omega }{\omega -W_{s}}\right) \right] ,\text{ }G_{b}^{0}(R_{i},R_{i},%
\omega )=\rho _{b}\left[ \ln \left( \frac{W_{b}+\omega }{\omega -W_{b}}%
\right) \right]   \label{flat}
\end{equation}
The self-consistent equations are rather easy to solve for this case and
allows us modify the parameters to fit the observed line shape. In Fig. 4 we
compare the analytical expression used by Knorr {\it et al.} \cite{knorr} to
fit the low energy part of $dI/dV$ and our results within the SBMFA. The
same set of parameters is used in section 8 to study the modifications of
the line shape in a circular corral. For the bulk density of states we take $%
\rho _{b}=0.145/$eV per site and spin from its value at the Fermi energy
reported by first-principles calculations \cite{moru}. $W_{b}$ is determined
from the filling of one electron per site $2W_{b}\rho _{b}=1$. From the
effective mass $m_{e}^{*}=0.38\ m_{e}$, where $m_{e}$ is the electron mass 
\cite{cro,euc} and a parabolic dispersion, one gets $\rho _{s}=0.045/$eV per
site and spin. From the bottom of the surface band we take  $W_{s}= 0.4$
eV, and we assume for simplicity the same value for the high energy cutoff.
As mentioned before, the results are only weakly sensitive to the cutoff. $%
E_{d}=-0.8$ eV is taken from Ref. \cite{uj}. The ratio $V_{b}/V_{s}$ is
determined by imposing a fixed ratio of the resonant level due to bulk ($\delta
_{b}=\pi \rho _{b}V_{b}^{2}$) or surface ($\delta _{s}=\pi \rho _{s}V_{s}^{2}
$) states: a) $\delta _{b}=\delta _{s}$, b) $\delta _{b}=3\delta _{s}$. The
magnitude of the hybridization controls the width of the line shape and is a
fitting parameter. The value of $p$ in Eq. (\ref{h}) is fixed in such a way
that for a clean surface, near 1/2 of the intensity of $dI/dV$ is due to
bulk states \cite{knorr,burg,jean}. Therefore we took $p=1/3$ to compensate
the approximate ratio $\rho _{b}/\rho _{s}\simeq 3$. Instead $q$ is used as
a fitting parameter which controls the asymmetry in $dI/dV$. In addition,
for the fit of Fig. 4, we shifted the minimum of $dI/dV$ and used a factor
that represents the quantity $\frac{4\pi e}{\hbar }A^{2}\rho _{t}$ in Eq. (%
\ref{i3}). From the fitting procedure we obtain a) $q=0.04$, $V_{s}=0.895$
eV and $V_{b}=0.499$ eV, b) $q=0.035$, $V_{s}=0.604$ eV and $V_{b}=0.583$ eV.

\vskip1.0cm

\begin{figure}[h!]
\hskip2.0cm\psfig{file=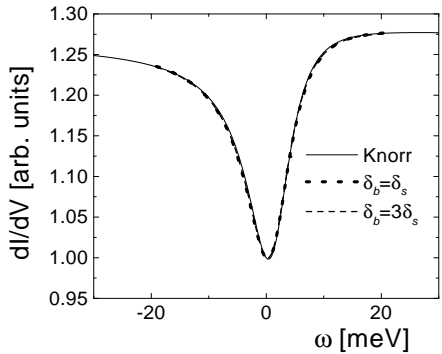,width=6.0cm,silent=} 
\caption{Comparison between the analytical expression used in Ref. 
17
to fit the low energy part of $dI/dV$ and our results within the
SBMFA. The parameters are explained in section 5 B.}
\end{figure}

\subsection{Exact diagonalization plus embedding (EDE)}

This method developed for impurity problems \cite{fer,buss}, consists in
solving numerically by the Lanczos method part of the system $H_{0}$ which
contains a finite number of relevant many-body states, and treating a
one-body term $H^{\prime }$ which connects it to the rest of the non
interacting system $H_{r}$, by an approximate method. For example, $H=H_{0}+$
$H^{\prime }+$ $H_{r}$ can describe a quantum wire with an embedded quantum
dot modeled by the impurity Anderson model in a chain \cite
{torio,torio2,wil2}. In this case $H_{0}$ contains the dot and a few
adjacent sites, and $H^{\prime }$ is the hopping of the extreme sites
included in $H_{0}$ to their nearest neighbors in $H_{r}$. For an impurity
in a quantum corral, $H_{0}$ should contain the impurity and a few
conduction eigenstates of the hard wall corral, which acquire a finite width 
$\delta $ due to hopping to the rest of the system \cite{wil}. As we show in
the next section, this width is essential to describe the physics.

The method starts by solving the one-particle Green's functions for $%
H^{\prime }=0$. Those for $H_{r}$ are known, and those of $H_{0}$ are
calculated using the recursion technique combined with the Lanczos method.
Off diagonal matrix elements are calculated from diagonal elements of hybrid
states, involving sum and difference of basis states. This information is
gathered in a matrix ${\bf g}$. For a non-interacting system ($U=0$), the
Green's function of $H$, which we denote by ${\bf G}$,
can be calculated from the Dyson equation ${\bf G=g+gH^{\prime }G}$. This is
taken as an approximation for the interacting system. Obviously the
approximation is exact for $H^{\prime }=0$ and any value of the interaction,
and also in the non-interacting case.

This approximation should be used with caution and incorrect results can be
obtained if it is applied outside its range of validity. For the Anderson
model, a reasonable criterion is that the size of the exactly solved part
should be smaller or of the order of the characteristic length $\xi \sim
\hbar v_{F}/T_{K}$ mentioned in section 4 \cite{torio,torio2,wil2}. In
practice, even when $\xi $ is ten times larger than the size of the system,
the resulting value of the impurity spectral density at the Fermi energy $%
\rho _{d\sigma }(\epsilon _{F})$ practically coincides with the exact value,
known from Friedel's sum rule \cite{torio,wil2} For much larger $\xi $, the
approximation is not valid. For example, the width of the Kondo resonance
near the symmetric case $E_{d}+U/2\sim \epsilon _{F}$ behaves as $U^{-2}$ 
\cite{wil2}, what is incorrect for large $U$ (implying small $T_{K}$ and
large $\xi $) \cite{note}. In the mirage experiment, using the velocity of
bulk states $v_{F}=1.57\times 10^{8}$ cm/s, then $\xi \sim 200$ nm\ and the
size of the ellipse is $2a\sim 14$ nm.

This technique is easier to implement than others for the case of more than
one impurity in the quantum corral and will be used in section 10.

\section{The role of confinement}

\subsection{One-body effects}

In the experiments of the mirage effect in an elliptical corral with
eccentricity $e=1/2$ and semimajor axis $a=7.13$ nm, the space dependence of 
$\Delta dI/dV$ reminds the wave function of the state number 42 for a
two-dimensional free electron gas in a hard-wall corral \cite{man}. This
already points out the importance of the confinement in the problem.
Although the hard wall is not a realistic assumption, some basic features of
the mirage effect can be understood with it \cite{wei,por}. The eigenstates
which determine the surface Green's function Eq. (\ref{g0}) have in general
a continuous distribution in energy, but the spectrum is discrete for a
hard-wall corral. From the form of the Schr\"{o}dinger equation we know that
for corrals of the same size, the separation between any two energy levels
is inversely proportional to the area of the corral \cite{lob1}. Therefore
in principle changing the size of the corral allows to single out one energy
level at will, place it near the Fermi energy $\epsilon _{F}$, and observe
it by STS, since as explained in section 4, it essentially captures the
conduction states near $\epsilon _{F}$. While this is a good starting point
for the understanding of the phenomenon, due to the soft character of the
walls, the corral eigenstates become resonances and there is a delicate
interplay between the width of these resonances and the separation between
energy levels.

While in presence of soft walls the surface eigenstates form a {\em continuum%
}, it turns out very useful not only for the understanding of the physics
but for the practical implementation of the many-body techniques, that under
general assumptions, the surface Green's function can be written as a {\em %
discrete} sum of contributions from resonances \cite{cal}. We have shown
this explicitly for the case of a circular confining potential of the form

\begin{equation}
V(r,\theta )=V_{\text{conf}}\frac{\hbar ^{2}}{2m_{e}^{*}r_{0}^{2}}\delta (%
\frac{r}{r_{0}}-1),  \label{pot}
\end{equation}
where $r,\theta $ are the polar coordinates on the plane and $V_{\text{conf}%
} $ is a dimensionless constant controlling the strength of the confinement
potential \cite{lob1,lob2}. The result is

\begin{equation}
G_{s}^{0}(r,\theta ,r^{\prime },\theta ^{\prime },\omega )=\sum_{n,m}\frac{%
C_{n}^{m}\ J_{m}(k_{n}^{m}r)J_{m}(k_{n}^{m}r^{\prime })\ e^{im(\theta
-\theta ^{\prime })}}{\omega -\epsilon _{n}^{m}+i\delta _{n}^{m}}.
\label{gop}
\end{equation}
The complex poles $\epsilon _{n}^{m}-i\delta _{n}^{m}=(\hbar
k_{n}^{m})^{2}/(2m_{e}^{*})$, where the complex wave vectors $k_{n}^{m}$ are
the zeros of a function $F_{m}(k)$ explained below which lead to positive $%
\delta _{n}^{m}$. The coefficients $C_{n}^{m}\ $are 
\begin{equation}
C_{n}^{m}=-\frac{i\ k_{n}^{m}}{\frac{\partial (F_{m}(k))}{\partial k}%
|_{k=k_{n}^{m}}}.  \label{res}
\end{equation}
$F_{m}(k)$ is a function of the Bessel functions of the first ($J_{m}$) and
second ($Y_{m}$) kind which is related to the normalization of the wave
functions in the continuum. Its expression is

\begin{eqnarray}
F_{m}(k) &=&A_{m}^{2}(k)+B_{m}^{2}(k)\text{, with }A_{m}(k)=1+\frac{V_{\text{%
conf}}(kr_{0})^{-1}}{\frac{Y_{m+1}(kr_{0})}{Y_{m}(kr_{0})}-\frac{%
J_{m+1}(kr_{0})}{J_{m}(kr_{0})}},  \nonumber \\
B_{m}(k) &=&\left( 1-A_{m}(k)\right) \frac{J_{m}(kr_{0})}{Y_{m}(kr_{0})}.
\label{ab}
\end{eqnarray}
In practice, in Eq.(\ref{gop}) one includes only the terms for which $%
\epsilon _{n}^{m}<E_{c}$, where the cutoff energy $E_{c}$ is of the order of
1 eV.

The results of Ref. \cite{lob1} suggest that for corrals of other shapes and
not too weak confinements, one can approximate the surface Green's function
as

\begin{equation}
G_{s}^{0}(R_{1},R_{2},\omega )\simeq \sum_{j}\frac{\varphi _{j}^{c}(R_{1})%
\overline{\varphi }_{j}^{c}(R_{2})}{\omega -\epsilon _{j}+i\delta _{j}},
\label{goap}
\end{equation}
where $\varphi _{j}^{c}(R)$ are the discrete eigenstates of the hard wall
corral and $\epsilon _{j}$ are their energies, calculated with a slightly
renormalized effective mass $m_{e}^{*}$ (increased by about 10\% \cite{lob1}), 
and $\delta _{j}$ are the width of the resonances, which to a very good
approximation are linear in energy $\delta _{j}=\delta _{F}(\epsilon
_{j}-\epsilon _{b})/(\epsilon _{F}-\epsilon _{b})$, where $\delta _{F}$ is
the width of the resonance at the Fermi level and $\epsilon _{b}$ is the
bottom of the surface band \cite{lob1}. As we show below, 
this width plays an essential
role in the many-body results. In some cases, for simplicity and since it
does not affect much the results, we will take constant $\delta _{j}=\delta
_{F}$.

\subsection{Many-body effects}

Usually, as in the simple case of section 3, the Anderson impurity is
hybridized with a continuous band of conduction states, flat on the scale of 
$T_{K}$. However, if one takes $V_{b}=0$ and a hard-wall assumption for the
surface states, the Anderson impurity in our model is mixed with a discrete
set of conduction states with a significant separation between adjacent
levels. Does a Kondo resonance form in this case? This question has been
addressed in the context of mesoscopic systems \cite{thi}. We illustrate it
with a simple problem of a ring of $L$ sites described by a tight binding
model with hopping $t$, in which one particular site has on-site energy 
$E_{d}=-t$, a Coulomb repulsion $U=2t$ and hopping $0.4t$ with their nearest
neighbors \cite{pc}. We take half filling, what implies $\epsilon _{F}=0$
and assume that the ring is threaded by half a flux quantum in order to
have (as in the case of the quantum mirage) an important hybridization 
of a conduction state at the Fermi energy 
$\epsilon _{F}$ with the impurity. This is a symmetric Anderson model 
with a discrete spectrum of conduction states. 
The impurity spectral density $\rho _{d\sigma }(\omega )$
calculated with perturbation theory in $U$ is shown in Fig. 5. For $L=800,$
the average separation between the levels which hybridize with the impurity $%
d=8t/L$ is an order of magnitude smaller than the half width of the
resonance $T_{K}\sim 0.1t$ and we can see a structure similar to Fig. 1 (a).
In particular, the Kondo resonance at $\epsilon _{F}$ can be visualized. For 
$L=80$ one has $d\sim T_{K}$ and the spectral function has some similarities
with that of the continuous conduction band, but with an important internal
structure. For $L=8$, for which $d\sim 10T_{K}$, the central Kondo peak is
absent \cite{note2}.

\begin{figure}[h!]
\hskip2.0cm\psfig{file=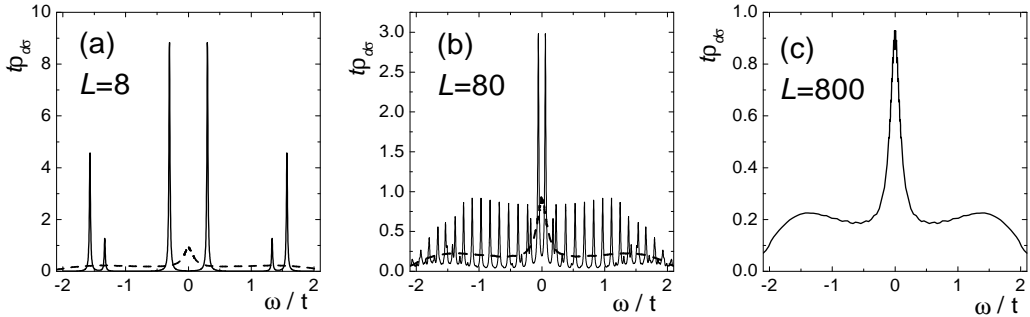,width=14.0cm,silent=} 
\vskip0.4cm
\caption{Impurity spectral density as a function of energy for a ring of 
$L$ sites described by the impurity Anderson model and several values of $L$.
An artificial broadening $\eta =0.01t$ was introduced for each peak. The
result for $L=800$ is also shown with a dashed line in (a) and (b) to
facilitate comparison. Parameters in the text (section 6 B).}
\end{figure}

As we see, to obtain a well defined Kondo resonance with discrete conduction
states, it is necessary that $d\lesssim T_{K}$. Instead, in the mirage
experiments one has $d\gg T_{K}$. While $T_{K}\sim 5$ meV \cite{man}, the
average distance between the energy levels that have an important
hybridization with the impurity (those shown in Fig. 10) is of the order of
100 meV. This shows the need to take into account the finite width $\delta
_{j}$ to the conduction states. The evolution with $\delta _{j}$ (taken for
simplicity independent of $j$) of $\rho _{d\sigma }(\omega )$ and the change
in the surface part of the conduction density of states after addition of
the impurity $\Delta \rho _{s\sigma }(\omega )$ for the elliptical corral
with $e=1/2$ studied experimentally \cite{man} is shown in Fig. 6. The
impurity is placed at one focus of the ellipse. $\rho _{s\sigma }(\omega )$
is given by Eqs. (\ref{h}) , (\ref{didv}) and (\ref{gh}) with $p=q=0.$ As
anticipated above, for very small $\delta $, the impurity spectral density
does not show a well defined resonance at $\epsilon _{F}$. As a consequence,
there is a marked disagreement of $\Delta \rho _{s\sigma }(\omega )$ with
the observed $\Delta dI/dV$ (which is very similar to the bottom left
curve). For $\delta =1$ meV, $\rho _{s\sigma }(\omega )$ has two peaks
(instead of antiresonances as in section 3) at the same positions of $\rho
_{d\sigma }(\omega )$, while in the absence of the impurity $\rho _{s\sigma
}(\omega )$ has a peak which corresponds to the state 42 which lies at $%
\epsilon _{F}$. Therefore the depression of $\Delta \rho _{s\sigma }(\omega
) $ at $\epsilon _{F}$ is a consequence of the subtraction and does not
indicate a Fano antiresonance. These results are consistent with numerical
results which correspond to $\delta =0$ but include (as usual in these
calculations) an artificial broadening of the resulting peaks \cite{hal}.
For a large broadening the results for $\Delta \rho _{s\sigma }(\omega )$
look like those of Fig. 6 for $\delta =10$ meV but with a large {\em positive%
} average, what is inconsistent with experiment.

\vskip2.0cm
\begin{figure}[h!]
\hskip2.0cm\psfig{file=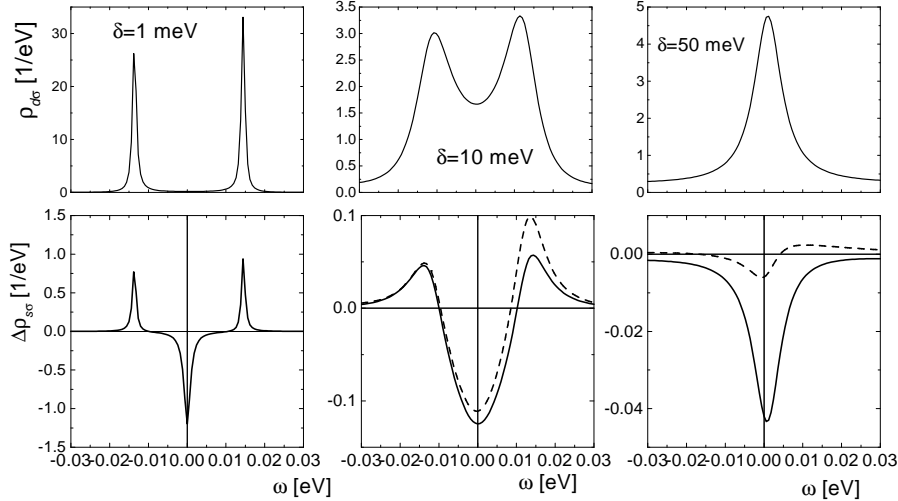,width=12.0cm,silent=} 
\vskip0.4cm
\caption{Impurity spectral density $\rho _{d\sigma }(\omega )$ (top) and
change in the surface spectral density due to addition of the impurity at
the impurity site $\Delta \rho _{s\sigma }(\omega )$ (bottom) for the
configuration of the mirage experiment and different values of the width of
the conduction levels $\delta :$ 1 meV (left), 10 meV (middle) and 50 meV
(right). The dashed line shows $\Delta \rho _{s\sigma }(\omega )$ at the
empty focus. Parameters as in Fig. 4 (section 5 A).}
\end{figure}

As $\delta $ increases, the two peaks in $\rho _{d\sigma }(\omega )$ merge
into one (for $\delta \sim 18$ meV) and the shape of both, the Kondo
resonance and the Fano antiresonance, become similar to the results of the
more conventional case, described qualitatively by the simple model of
section 3. The Fano dip in $\Delta \rho _{s\sigma }(\omega )$ for $\delta
=50 $ meV agrees well with experiment \cite{man}. The rather symmetrical
shape is due to the fact that $V_{s}$ decreasing with energy was assumed.
For constant $V_{s}$, $\Delta \rho _{s\sigma }(\omega )$ is smaller for
positive $\omega $ (see full line of Fig. 16). 
The evolution of $\rho _{d\sigma }(\omega )$ with $%
\delta $, described first in Ref. \cite{rap}, has been confirmed by exact
diagonalization plus embedding \cite{wil}, and by Wilson renormalization
group calculations \cite{cor1}.

In Fig. 6 we also show $\Delta \rho _{s\sigma }(\omega )$ at the empty
focus. The comparison with the corresponding value at the focus where the
impurity is located establishes the intensity of the mirage effect. For
small $\delta $ the ``transmission'' of the Kondo effect to the empty focus
is nearly perfect, because the space dependence follows closely the density
of the state 42 which has maxima at the foci (see Fig. 8). As $\delta $
increases, the intensity of the mirage decreases as a consequence of
interference effects described in the next section.

\vskip3.0cm
\begin{figure}[h!]
\hskip2.0cm\psfig{file=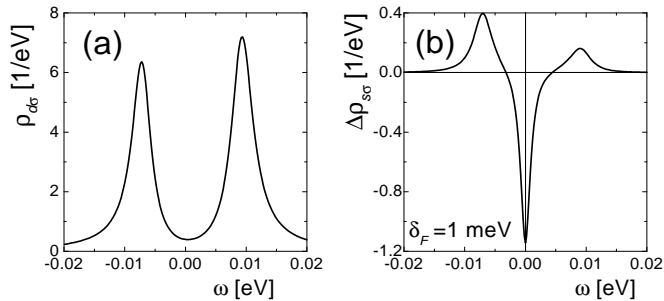,width=9.0cm,silent=} 
\vskip0.4cm
\caption{(a) Impurity spectral density and (b) $\Delta \rho _{s\sigma
}(\omega )$ at the impurity site as a function of energy for $\delta
_{j}=\delta _{F}(\epsilon _{j}-\epsilon _{b})/(\epsilon _{F}-\epsilon _{b})$
with $\delta _{F}=1$ meV. Other parameters are $V_{s}=0.45$ eV, $\delta
_{b}=35$ meV, $E_{d}^{eff}=7$ meV, and $U=1$ eV.}
\end{figure}

The introduction of moderate hybridization of the impurity with bulk states
does not affect the need to include a non vanishing $\delta $ to be able to
obtain a reasonable agreement with experiment. In Fig. 7 we show again both
densities for $\delta =1$ meV and parameters such that the strength of the
hybridization of the impurity with bulk and surface states is approximately
the same \cite{note4}. The peaks in $\rho _{d\sigma }(\omega )$ and $\rho
_{s\sigma }(\omega )$ are broadened with respect to the previous case, but
again the dip in $\Delta \rho _{s\sigma }(\omega )$ is not a Fano
antiresonance, but correspond to minus the peak in $\rho _{s\sigma }(\omega
) $ at $\epsilon _{F}$ in absence of the impurity.

\section{The space dependence of $dI/dV$}

While scattering theories based on a phenomenological phase shift for the
scattering at the atoms of the boundary and the impurity describe
quantitatively the space dependence \cite{fiete,agam,fie}, approaches based
on wave functions of a corral (with continuous boundaries) usually bring
more insight into the underlying physics \cite{wei,por,rap}. For example the
prediction of mirages out of the foci of elliptical corrals are somewhat
hidden in the scattering approaches. Instead, mirages observed in a circular
corral \cite{man2} were inspired by the extrema of the wave functions of the
degenerate $37^{th}$ and $38^{th}$ conduction eigenstates of a hard-wall
circular corral, and were calculated with our many-body approach for a
circular corral with soft walls \cite{lob1,lob2}.

\begin{figure}[h!]
\hskip2.0cm\psfig{file=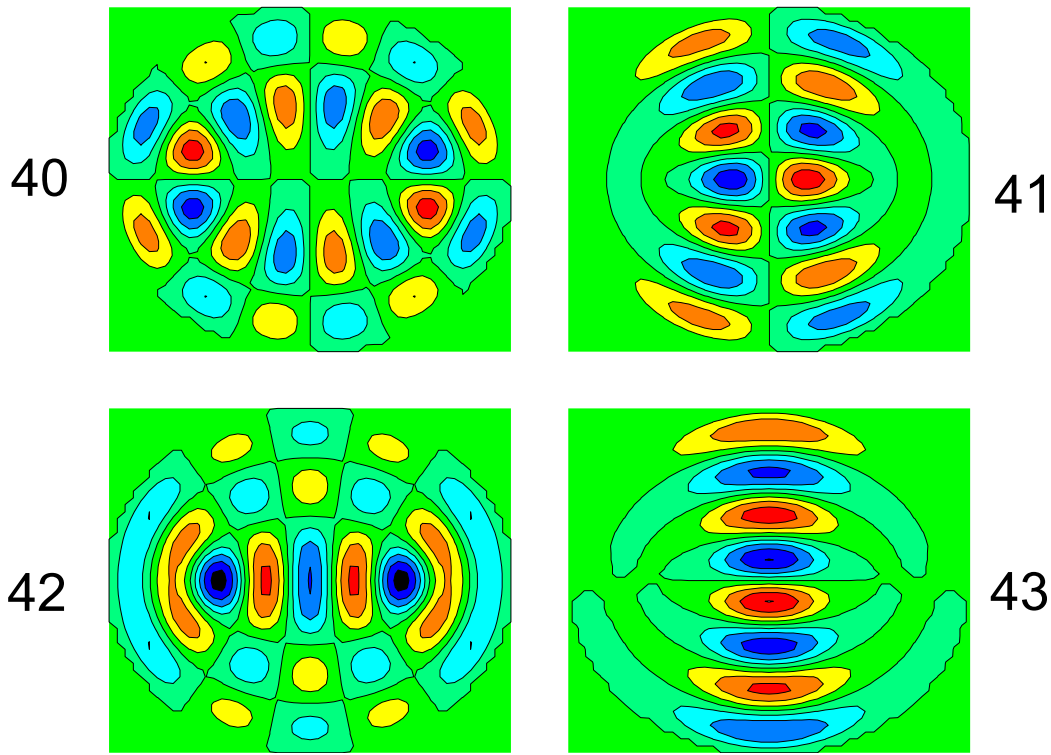,width=14.0cm,silent=} 
\vskip0.4cm
\caption{Contour plot of the wave functions of an elliptical corral with
semimajor axis $a=7.13$ nm and eccentricity $e=1/2$ which lie close to the
Fermi energy.}
\end{figure}

\vskip2.0cm

\begin{figure}[h!]
\hskip2.0cm\psfig{file=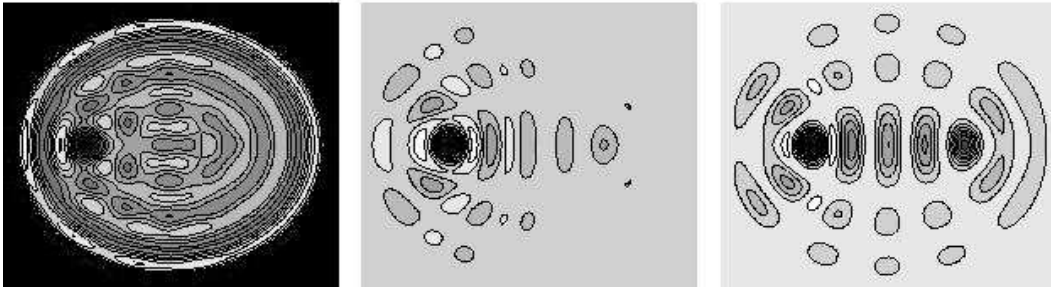,width=14.0cm,silent=} 
\vskip0.4cm
\caption{{Contour plot of }$dI/dV$\ {in the elliptical corral with } $%
a=7.13$ nm and $e=1/2$ {for $\delta =50$ meV (left), $\Delta dI/dV$ for ${%
\delta =50}$ {meV {(middle)}, and }$\Delta dI/dV$ for $\delta =20$ meV {%
(right). The applied voltage is 10 mV. }}Other parameters as in Fig. 4
(section 5 A).}
\end{figure}

Having in mind the most studied case of the mirage effect: a Co impurity
placed at one focus of an elliptical corral with $e=1/2$ built on a Cu(111)
surface \cite{man}, we have calculated the differential conductance 
$dI/dV(R_{t},R_{i},V)$ as a function of the tip position $R_{t}$, for the
impurity position fixed at the left focus [$R_{i}=(-0.5a,0)$] and voltage $V=10$ mV. 
We used Eqs. (\ref{didv}), (\ref{gh}) and (\ref{goap}). We have
taken $p=q=0$, since they are important only near the impurity \cite
{note5}. Therefore the results depend on the impurity Green's function $%
G_{d}(\omega )$ calculated as is section 4 A, and mainly on the unperturbed
surface Green's function $G_{s}^{0}(R_{1},R_{2},\omega )$. To calculate the
latter we used Eq. (\ref{goap}) with the corral wave functions $\varphi
_{j}^{c}(R)$ calculated as in Ref. \cite{nakam}. The wave functions of the
states which lie nearer to the Fermi energy are shown in Fig. 8. The wave
functions can be classified by symmetry into the four irreducible
representations of the point group $C_{2v}$, according to the parity under
reflection through the major (minor) axis $\sigma _{y}$ ($\sigma _{x}$). In
particular each of the shown wave functions belongs to a different
representation. The state 42, which lies at the Fermi energy is even under
both reflections, 40 is odd under both of them, 41 is even under $\sigma
_{y} $ and odd under $\sigma _{x}$, and 43 is odd under $\sigma _{y}$ and
even under $\sigma _{x}$.

The results presented in Fig. 9 were obtained for $V_{b}=0$, but quite
similar results come out if $V_{s}$ is decreased by a
factor $1/\sqrt{2}$ and $V_{b}$ is increased 
so that the contribution to the resonant level width of
bulk and surface states has the same magnitude, and the width of the impurity
spectral density is kept. This is not surprising since the above mentioned
change of parameters practically does not affect $G_{d}(\omega )$, and then,
from Eq. (\ref{gh}), the only change in $\Delta dI/dV(R_{t})$ for $p=q=0$
comes from a factor $V_{s}^{2}$ (see Fig. 3 of Ref. \cite{lob1}). Instead,
the dependence on the impurity position $R_{i}$ should be affected by the
relative strength of $V_{b}$ and $V_{s}$ (see next section).

The differential conductance $dI/dV(R_{t})$ for a constant width $\delta =50$
meV of the conduction surface states is represented in Fig. 9 left. It is
very similar to the observed topograph \cite{man}. However, the latter
corresponds to the total current $I$ and not to $dI/dV$. The similarity is
due to the fact that $\delta $ is larger than the energy corresponding to
the applied voltage $eV=10$ meV, and $dI/dV$ does not change too much in this
energy scale. Comparing with Fig. 8, one sees that as a first approximation,
the observed pattern can be described as a sum of the densities of the state
42 which lies at the Fermi energy $\epsilon _{F}$, and the state 43 which is 
$\sim eV$ above $\epsilon _{F}$. The wave function of the state 42 shows
some vertical ``stripes'' which end in ``arcs'' at the extreme left and
right. These essential features rotated 90 degrees describe roughly the wave
function of the state 43. Therefore the structure with ``arcs'' at the
border and ``stripes'' in the middle is to be expected in the sum of
probability densities.

Translated into equations, this is consistent with the behavior expected
from the first term of the first Eq. (\ref{gh}) and Eq. (\ref{goap}).
However, this is not the whole story because the above mentioned term
depends on the sum of squares of wave functions and is therefore invariant
under all symmetry operations of the ellipse, while the observed topograph
(and our calculated $dI/dV$) has not a defined parity under $\sigma _{x}$.
The addition of the impurity breaks the symmetry under $\sigma _{x}$ and the
effect of the impurity is contained completely in $\Delta G_{h}$ [see Eq. (%
\ref{gh})]. The imaginary part of $\Delta G_{h}$ is directly proportional to 
$\Delta dI/dV$, which is $dI/dV$ minus the corresponding quantity for the
empty corral. From Eqs. (\ref{gh}) and Eq. (\ref{goap}), it is clear that
physically the effect of this subtraction is to eliminate the contribution
of all states $j$ which have a negligible hybridization with the impurity $%
V_{s}\varphi _{j}^{c}(R_{i})$. In particular, states odd under $\sigma _{y}$
like 40 and 43 have $\varphi _{j}^{c}(R_{i})=0$ and do not participate in $%
\Delta dI/dV$. In practice, states like 41 which are even under $\sigma _{y}$
but have a small amplitude at the foci do not affect the result either. In
Fig. 9 middle we show the ``cleaned'' result $\Delta dI/dV$ for the same
parameters of the complete result $dI/dV$ shown at the left. Now the main
features of the wave function of the state 42 can be recognized directly,
particularly if the width of the conduction levels is reduced to $\delta =20$
meV (Fig. 9 right). Comparison with experiment \cite{man} indicates that the
right value of $\delta $ is in between those shown: $20$ meV $<\delta <50$
meV.

\begin{figure}[h!]
\hskip2.0cm\psfig{file=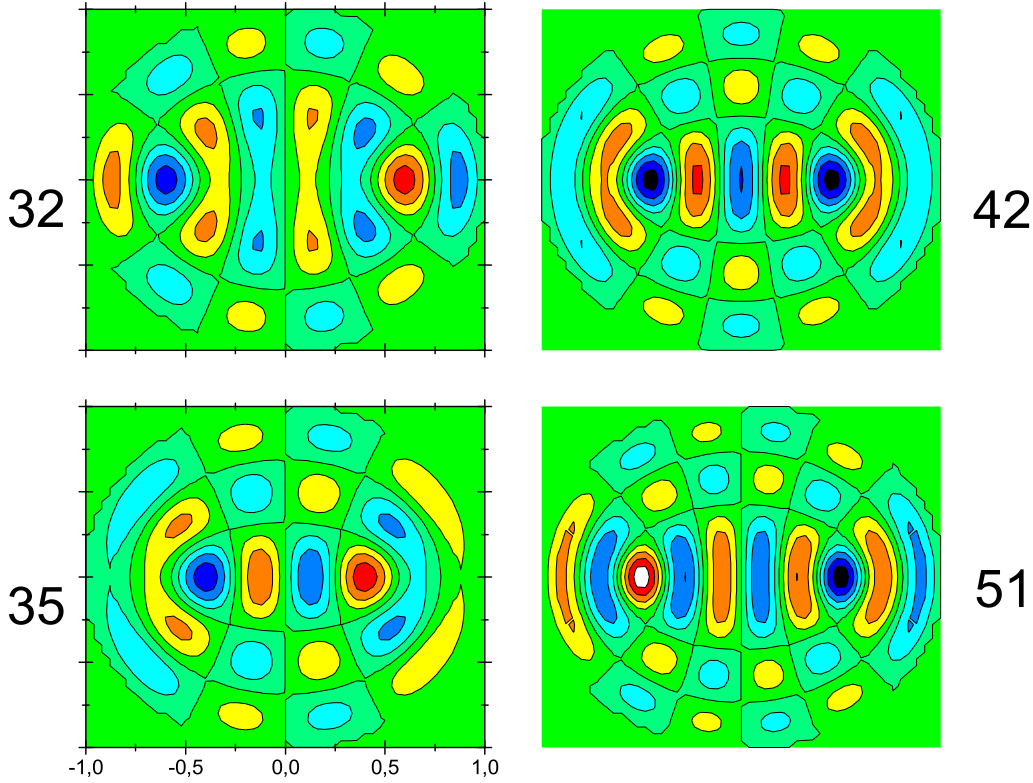,width=14.0cm,silent=} 
\vskip0.4cm
\caption{Contour plot of the wave functions of an elliptical corral
with semimajor axis $a=7.13$ nm and $e=1/2$ with appreciable amplitude and
the foci and close to the Fermi energy.}
\end{figure}

An analysis of the magnitude of the wave functions at the foci (which
determine the hybridization strength of the impurity with the different
states) shows that the space dependence of $\Delta dI/dV$ is dominated by
four states. The wave functions of these states are shown in Fig. 10. While
all these states are even under $\sigma _{y}$ (otherwise they would not
hybridize with the impurity), only 42 is even under $\sigma _{x}$. The rest
are odd under $\sigma _{x}$. This produces a negative interference between
the contribution of the state 42 and the other three at the empty focus
which tends to destroy the mirage effect. In simple terms, one could say
that the information of the Kondo effect transmitted by the focus of the
impurity by the four wave functions reaches the other focus with positive
sign for the states 42 and with negative sign by the states 32, 35 and 51 so
that the amplitude is reduced. Formally, this can be seen in Eqs. (\ref{gh})
and (\ref{goap}). As $\delta $ decreases, the relative contribution of the
state 42 which lies at the Fermi energy increases and the size of the mirage
effect also increases. This suggest to reduce $\delta $, if one can control
this parameter experimentally, or to try to optimize the geometry 
in order to reduce
the negative interference effects \cite{rap}. However, as shown in the
previous section, if $\delta $ is reduced too much, the Kondo resonance and
Fano antiresonance near the Fermi level are destroyed.

\section{Voltage dependence of $\Delta dI/dV$}

The experimental study of the line shape of $\Delta dI/dV$ at the impurity
site in different positions of one corral or in different corrals and its
comparison with theory should be useful to elucidate the relative strength
of the hybridization of the magnetic impurity with bulk and surface states.
A greater sensitivity to geometry points towards a greater relevance of
surface states. Recently, it has been argued that due to the exponential
dependence of the Kondo temperature on the density of states \cite{note},
the observed line shape with approximately the same width in different
situations, indicates that the hybridization with bulk states should be much
more important \cite{cor2}. However, the calculations of Ref. \cite{cor2}
are rather generic and the specific features of the corral states were not
taken into account.

\vskip5.0cm
\begin{figure}[h!]
\hskip2.0cm\psfig{file=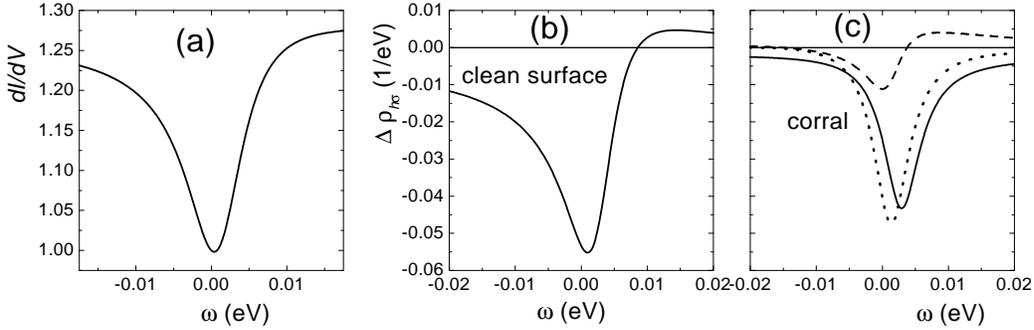,width=14.0cm,silent=} 
\vskip0.4cm
\caption{(a) Analytical function used to fit $dI/dV$
for a Co atom on a clean Cu(111) surface 
(Ref. 17).
(b) Density of the 
$h_{\sigma }$ state proportional to $\Delta dI/dV$ 
[see Eq. (11)]
for
the system of (a). (c) same as (b) for a Co atom at the focus of the
elliptical corral of Fig. 10, at the Co position (full line) and at the
empty focus (dashed line). For comparison, the dotted line is the result for
the clean surface with $q=0$. Parameters are $V_{s}=0.64$ eV, $\delta =40$
meV, $U=1$ eV and $q=0.03$.}
\end{figure}

For the case of the Cu(111) surface, to the best of our knowledge the
dependence of $\Delta dI/dV$ on bias voltage has been reported only in two
cases: the clean surface \cite{man,knorr} (see Fig. 4) and the elliptical
corral described in the previous section, with a Co atom at one of the foci 
\cite{man}. In the latter case the line shape is more symmetric, but the
width is approximately the same in both cases. Both line shapes can be
qualitatively described  including {\em only }
hybridization with surface states. In Fig. 11 we show our results obtained
within perturbation theory, using Eqs. (\ref{gh}), with $p=0$, and $q=0.03$,
to control the asymmetry of the line. Since we used here constant $V_{s}$
and surface density of states $\rho _{s}$ (as in section 3), and the nearly
symmetric case $E_{d}+U/2\sim \epsilon _{F}$, the line shape for the clean
surface is symmetric for $q=0$ (dotted line in Fig. 11 (c)), and a value of $%
q>0$ reproduces the observed asymmetry (Fig. 11 (a)). Instead, a constant $%
V_{s}$ in the corral case leads to an asymmetry {\em opposite }to that
observed for the clean surface (like the full line of Fig. 16), while the
line shape observed in the corral is symmetric. Then, in this case the
effect of $q>0$ is to correct the asymmetry. It is encouraging that the same
set of reasonable parameters can explain qualitatively both line shapes. The
experimental $\Delta dI/dV$ has kinks around $\pm 0.01$ V which are probably
due to peculiarities of the non-interacting band structure and are out of
the scope of our theory \cite{man,knorr}. As shown at the bottom of Fig. 6,
the width of $\Delta dI/dV$ has some variation with the width of the
conduction states $\delta $. Here we have chosen $\delta =40$ meV, which as
discussed in the previous section, leads to a space dependence of $dI/dV$ in
agreement with experiment.

\vskip2.0cm
\begin{figure}[h!]
\hskip2.0cm\psfig{file=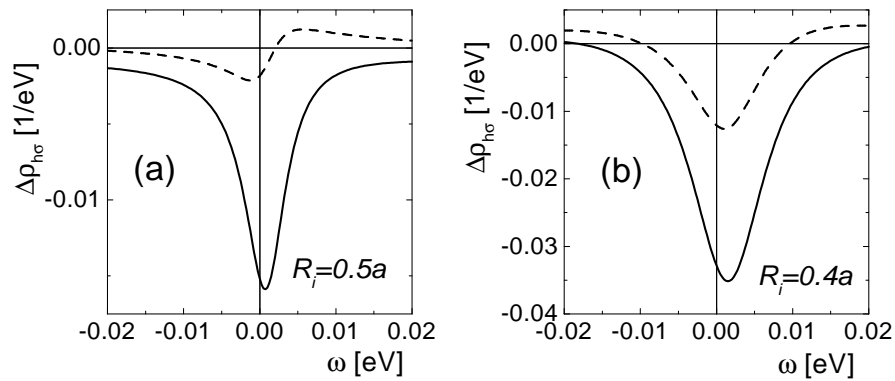,width=12.0cm,silent=} 
\vskip0.4cm
\caption {$\Delta \rho _{h \sigma} (\omega ) \sim \Delta dI/dV(\omega
/e)$ 
[see Eq. (11)] 
for an ellipse with $a=6.46$ nm and $e=1/2$, and
two impurity positions: (a) $R_{i}$ at one focus, (b) $R_{i}=(\pm 0.4a,0)$.
Full (dashed) line corresponds to the tip at $R_{i}$ ($-R_{i}$). Parameters
are $\delta _j = \delta _F(\epsilon _j - \epsilon _b)/(\epsilon
_F-\epsilon _b)$ with $\delta _F = 40$ meV, $V_s=0.48$ eV, $\delta
_b=32$ meV, $E_d^{eff}=7$ meV, $U=1$ eV, $p=0$ and $q=0.02$.}
\end{figure}

The rather similar line widths in the above mentioned cases seems accidental
and other situations are more suitable to analyze the relative role of
surface and bulk states in the formation of the Kondo resonance. In Fig. 12
we show the line shape expected in a smaller elliptical corral, with
semimajor axis reduced to $a=6.46$ nm keeping the same eccentricity $e=1/2$,
so that the state 35 (see Fig. 10) falls at the Fermi energy. This state has
extrema at positions $R_{e}=(\pm 0.4a,0)$, and the average separation of the
levels is larger than in the previous case. The surface spectral density
near the Fermi level is larger at $R_{e}$ than at the foci $(\pm 0.5a,0)$.
Even including the same hybridization strength of the impurity with surface
and bulk states, the depth and width of $\Delta dI/dV$ is substantially
larger if the impurity is placed at $R_{e}$ rather than at the foci. Note
also that the intensity with the tip at the opposite point $R_{t}=-R_{i}$ is
considerable larger in this case. This is due to the fact that the negative
interference between states 42 and 35 explained in the previous case is
substantially reduced. A stronger mirage in this geometry has been predicted
before \cite{rap}.

In the rest of this section, we show results for the line shape for the tip
placed on the impurity and several positions of the impurity inside a
circular corral of radius $r_{0}=6.35$ nm, in such a way that the degenerate
states 37 and 38 lie at the Fermi level. Experiments in this corral have
been done to illustrate the simultaneous presence of two mirages \cite{man2}%
. We use the SBMFA, because it gives the correct exponential dependence of $%
T_{K}$ with the coupling constant \cite{note}. We also use the exact Eqs. (%
\ref{gop}) to (\ref{ab}) for the surface Green's function instead of the
approximate one Eq. (\ref{goap}). The SBMFA is described in section 5 B and
the parameters are those taken there to fit the line shape for the clean
surface.

\vskip3.0cm
\begin{figure}[t!]
\hskip2.0cm\psfig{file=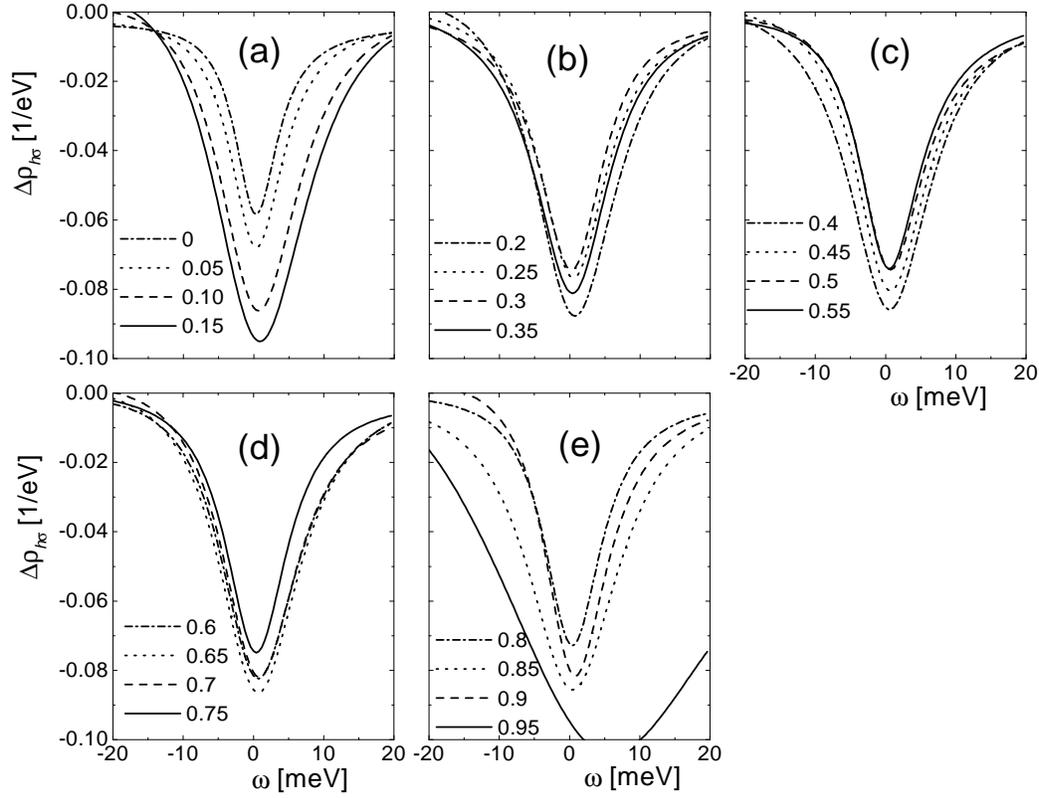,width=14.0cm,silent=} 
\vskip0.4cm
\caption{$\Delta \rho _{h\sigma} (\omega )\sim \Delta dI/dV(\omega
/e)$ within SBMFA for a circular corral of radius $r_{0}=6.35$ nm at the
impurity position ($R_{t}=R_{i}$), for $\delta _{F}=40$ meV, $\delta
_{b}=\delta _{s}$ and different values of $|R_{i}|/r_{0}$ indicated inside
each figure. Other parameters as in Fig. 4 (section 5 B).}
\end{figure}

In absence of the impurity, the density of surface conduction electrons at
the Fermi energy has a pronounced relative maximum near $r=|R_{i}|=0.15r_{0}$
\cite{lob1}. As shown in Fig. 13 (a), the depth and width of $\Delta dI/dV$
varies considerably as the impurity and STM\ tip are moved together from the
center of the corral to this maximum. At the center, only the corral surface
states with angular momentum projection $m=0$ can hybridize with the
impurity. Since the corresponding resonances are far from the Fermi level
(see Fig. 14), the Fano
antiresonance for $r=0$ is more than 80\% due to bulk states. In fact, doing
the same calculation with $p=0$ (assuming no hopping between tip and bulk
states, see Eqs. (\ref{h}) and (\ref{gh})), $|\Delta \rho _{h\sigma }|<0.01/$%
eV. Therefore for $r=0$ the bulk states play a major role not only in the
formation of the Kondo state but also in the variation of the STM current $%
\Delta dI/dV$ which is mainly {\em due to the current between tip and bulk
states}. At remote positions this Fano antiresonance of bulk states will not
be observed \cite{note5}. In general, the contribution to the dip in $\Delta
dI/dV$ due to bulk states (and interference with surface states), captured
at the impurity by the hybridization of tip and bulk states, will be absent
at a mirage point and is a natural limitation of the intensity at the mirage
point (see next section). For $r/r_{0}=0.15$, the intensity of $\Delta dI/dV$
decreases to $\sim $ 40\% if the tip-bulk hopping is disconnected.

Compared with the rapid variation for $r/r_{0}<0.2$, the width and
magnitude of $\Delta dI/dV$
oscillates weakly with position for $0.2<$ $r/r_{0}<0.9$, with larger
intensity and width for $r/r_{0}=0.4$, 0.65, and 0.85 (see Fig. 13).
However, there is a dramatic increase for $r/r_{0}>0.9$, with a maximum near
0.96, as shown in Fig. 13 (e). Although unfortunately at this short
distances from the boundary our theory ceases to be reliable (because of our
simple assumption of a continuous boundary potential), it is instructive to
relate this result with the variation of the density of surface states $\rho
_{s\sigma }^{0}(r,\omega )$ at the Fermi energy $\epsilon _{F}=0$ with
position. As shown in Fig. 14, there is a moderate increase in $\rho
_{s\sigma }^{0}(r,\omega )$ near $\omega =0$ as $r/r_{0}$ increases from
0.15 to 0.95. This is mainly due to the contribution of resonances with high
angular momentum projection which render $\rho _{s\sigma }^{0}$ rather flat
in energy. Instead, for $r/r_{0}=0$, the resonances with $m=0$ are selected
and they lead to the displayed oscillatory behavior.

\vskip2.0cm
\begin{figure}[h!]
\hskip2.0cm\psfig{file=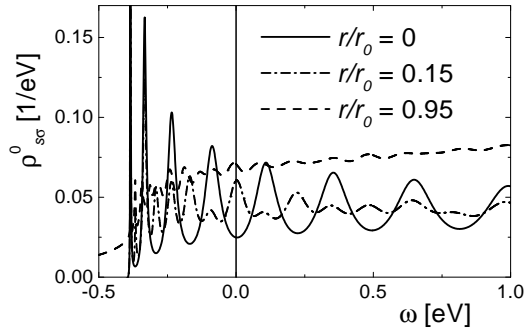,width=7.0cm,silent=} 
\vskip0.4cm
\caption{Density of unperturbed surface states $\rho _{s\sigma
}^0(r,\omega )=-\lambda ^2 {\rm Im}
[G_{s}^{0}(r,\theta ,r,\theta ,\omega )/\pi]$ [see Eqs. (\ref{gop}) to 
(\ref{ab})] as a function of energy $\omega $ for different values of $r/r_{0}$.}
\end{figure}

The parameters of Fig. 13 correspond to an equal participation of bulk and
surface states in the resonant level $\delta _{b}=\delta _{s}$. Considering
the case $\delta _{b}=3\delta _{s}$, as expected, the variation of the width
of the resonance with the position of the impurity is less pronounced, but
otherwise the same qualitative features as before are obtained. Except for
the peculiar behavior near the boundary of the corral, the greater
sensitivity to the position is for $0<$ $r/r_{0}<0.15$ as before. Comparison
between both cases is presented in Fig. 15. In Fig. 15 (a) and (b) we have
used an intensity of the boundary potential $V_{\text{conf}}=7$, which leads
to a broadening $\delta _{F}=40$ meV of the surface conduction states at the
Fermi level (see section 6 A). This value leads to a space dependence in
agreement with experiment (see section 7). In Fig. 15 (c) we show how the
space variation is affected if the confining potential is increased to $V_{%
\text{conf}}=15$, leading to $\delta _{F}=20$ meV. The oscillations in the
surface density of states and therefore the variation of the width of the
resonance with position becomes much more pronounced. There is a tendency to
a change in the line shape, similar to that of Fig. 6 for $\delta _{F}=10$
meV, which can allow to identify or rule out this regime experimentally. For
other positions of the impurity (not shown), the tendency is similar to that
of Fig. 13, but the particular structure for $r/r_{0}=0.95$ almost
disappeared.

\vskip2.0cm
\begin{figure}[h!]
\hskip2.0cm\psfig{file=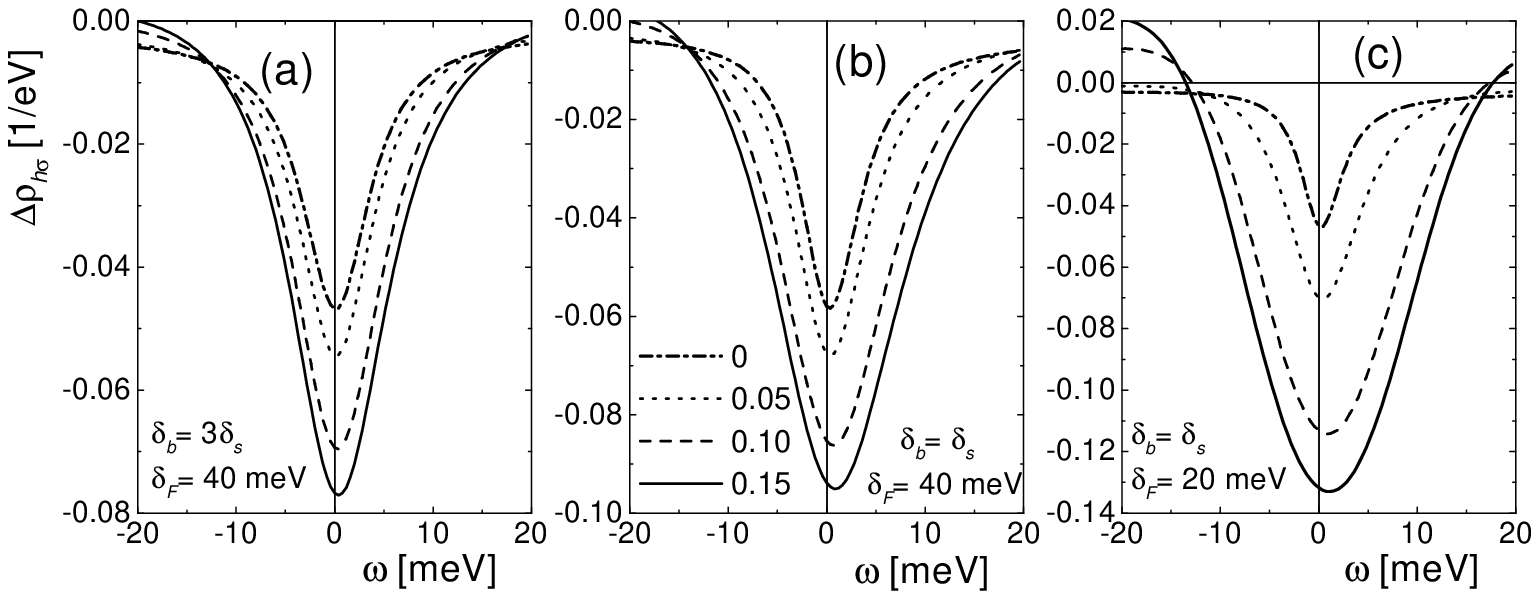,width=14.0cm,silent=} 
\vskip0.4cm
\caption{Same as Fig. 13 for (a) $\delta _F=40$ meV, $\delta
_{b}=3\delta _s$, (b) $\delta _F=40$ meV, $\delta _b=\delta _s$ and
(c) $\delta _F=20$ meV, $\delta _b=\delta _s$.}
\end{figure}

\section{Lower bound for impurity-surface hybridization}

Within our local picture for the hybridization of the impurity and tip with
conduction states, a simple estimate of a lower bound for $V_{s}/V_{b}$ can
be obtained from the mirage experiment in the elliptical quantum corral \cite
{man}. The ratio of the intensity of $\Delta dI/dV$ at the mirage
point $I_{m}$ (for the tip position $R_{t}=-R_{i}$)  to that at the impurity 
$I_{i}$ (for $R_{t}=R_{i}$) was reported to be $I_{m}/I_{i}\cong 1/8$. As in
section 3, let us approximate $G_{b}^{0}(R_{i},R_{i},\omega )\simeq -i\pi
\rho _{b}$. Also, for enough broadening of the surface conduction states 
$\delta \geq 40$ meV one has $G_{s}^{0}(R_{i},R_{i},\omega )\simeq -i\pi \rho
_{s}$ \cite{rap}. Neglecting the tip-impurity hopping ($q=0$), using Eqs. (\ref{didv})
and (\ref{gh}) one has 
\begin{equation}
I_{i}=-\Delta dI/dV(R_{t})\simeq C\left( \frac{V_{s}}{V_{b}}+p\frac{\rho _{b}%
}{\rho _{s}}\right) ^{2}\rho _{d}(\omega )=C\left( \frac{V_{s}}{V_{b}}%
+1\right) ^{2}\rho _{d}(\omega ),  \label{ii}
\end{equation}
where $C=\pi V_{b}^{2}\rho _{s}^{2}$ is a constant and in the last equality
we assumed that $p\rho _{b}=\rho _{s}$, so that in the absence of impurity,
the tip detects bulk and surface states with the same intensity, as reported
experimentally \cite{knorr,burg,jean}. Now, at the mirage point one can
neglect $G_{b}^{0}(R_{i},-R_{i},\omega )$ \cite{note5}. Assuming instead
perfect transmission from the surface states (as if only the state 42 were
relevant), one has $G_{s}^{0}(R_{i},-R_{i},\omega
)=G_{s}^{0}(R_{i},R_{i},\omega )$. Since the amplitude at the mirage point
is less than that for perfect transmission, one has from Eqs. (\ref{didv})
and (\ref{gh})

\begin{equation}
I_{m}=-\Delta dI/dV(-R_{t})<C\left( \frac{V_{s}}{V_{b}}\right) ^{2}\rho
_{d}(\omega ),  \label{im}
\end{equation}
and then, from Eqs. (\ref{ii}) and (\ref{im})

\begin{equation}
\frac{V_{s}}{V_{b}}>\sqrt{\frac{I_{m}}{I_{i}}}\left( 1+\frac{V_{s}}{V_{b}}%
\right) >\sqrt{\frac{I_{m}}{I_{i}}}+\frac{I_{m}}{I_{i}}\text{.}
\label{bound}
\end{equation}
Solving a quadratic equation, a more precise bound for $I_{m}/I_{i}=1/8$
gives $V_{s}/V_{b}>0.547$. Using $\rho _{b}\simeq 3\rho _{s}$, this implies $%
\delta _{s}>\delta _{b}/10$, with $\delta _{c}=\pi \rho _{c}V_{c}^{2}$. A
smaller tip-bulk hopping $p$ leads to a larger lower bound for $\delta _{s}$.

\section{Interaction between Kondo impurities in a quantum corral}

Experiments with two impurities inside an elliptical quantum corral have
been done \cite{man2}, but the results were not published yet. These
experiments should be particularly useful as a test of the relative strength
of the hybridization of the impurity with bulk and surface states, since one
expects that at distances larger than $\sim $0.5 nm, the interaction between
two Kondo impurities is dominated by surface states. In this section we
extend previous calculations of the line shape of $\Delta dI/dV$ when there
is one impurity at each focus of an elliptical corral, using the technique
of exact diagonalization plus embedding, described in section 5 C \cite{wil}.
Other calculations of the interaction between magnetic impurities in a
corral have been made by perturbation theory in the Kondo coupling \cite
{correa}. However, this technique does not work in the case we are
interested of antiferromagnetic Kondo coupling $J_{K}>0$ \cite{note}.

As explained in section 5 C, the Hamiltonian is written as $H=H_{0}+$ $%
H^{\prime }+$ $H_{r}$. In our case the Hilbert space of $H_{0}$ contains one
or two impurities and the most important surface conduction states (those
represented in Fig. 10) and two additional ones (24 and 62) although they do
not affect the results. $H_{r}$ describes a set of independent non-interacting bulk
states which hybridize independently with the impurities and the surface
conduction states. $H_{0}$ has a similar form to Eq. (\ref{ham}) but only
contains hard-wall surface conduction states and can contain more than one impurity:

\begin{equation}
H_{0}=\sum_{j\sigma }\epsilon _{j}s_{j\sigma }^{\dagger }s_{j\sigma
}+E_{d}\sum_{i\sigma }d_{i\sigma }^{\dagger }d_{i\sigma
}+U\sum_{i}d_{i\uparrow }^{\dagger }d_{i\uparrow }d_{i\downarrow }^{\dagger
}d_{i\downarrow }+\sum_{ij\sigma }\lambda V_{s}[\varphi
_{j}(R_{i})d_{i\sigma }^{\dagger }s_{j\sigma }+\text{H.c.}],  \label{h0}
\end{equation}
while the effect of bulk states and the broadening of the surface states
(necessary to obtain a qualitatively reasonable line shape as shown in
section 6) is contained in $H^{\prime }$ which reads

\begin{equation}
H^{\prime }\cong t\sum_{j\sigma }(s_{j\sigma }^{\dagger }b_{j\sigma }+\text{%
H.c.})+V_{b}\sum_{i\sigma }(d_{i\sigma }^{\dagger }b_{i\sigma }+\text{H.c.}).
\label{h2}
\end{equation}
For the bulk states $b_{l\sigma }$ we take a constant unperturbed density of 
$\rho _{b}=0.05$ states/eV per spin (of the order of the density of bulk $s$
and $p$ states at $\epsilon _{F}$.\cite{euc}), but a change in $\rho _{b}$
can be absorbed in a change in $t$ and $V_{b}$. The value of $t$ controls
the width of the conduction states and therefore, the intensity at the
mirage point, as explained in sections 6 and 7.

\vskip3.0cm
\begin{figure}[h!]
\hskip2.0cm\psfig{file=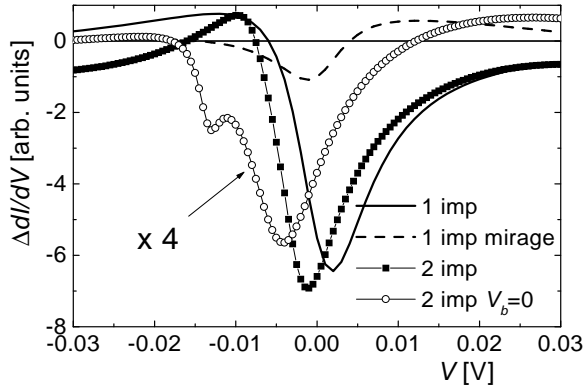,width=8.0cm,silent=} 
\vskip0.4cm
\caption{$\Delta dI/dV$ as a function of applied voltage calculated
with the EDE for an elliptical corral with one or two impurities at the foci.
Full line: one impurity and tip on it. Dashed line: one impurity and tip at
the other foci. Solid squares: two impurities and tip on one of them. Open
circles: same as before with $V_s$ increased by a factor $\sqrt{2}$ and $%
V_{b}=0$. Unless otherwise indicated parameters are $E_d=-0.8$ eV, $U=3$
eV 
(from Ref. 67), 
$V_{s}=1.12/\sqrt{2}$ eV, $V_{b}=1.2$ eV, $p=q=0$.}
\end{figure}

In Fig. 16, we represent the change in differential conductance for $p=q=0$
for the case in which there is one impurity at each focus, comparing two
situations. In the first one $V_{b}=0$ and $V_{s}=1.12$ eV is taken to
reproduce approximately the experimental width. In the second one $V_{s}$ is
reduced by a factor $1/\sqrt{2}$ and $V_{b}$ increased to 1.2 eV in order
that the same width and practically the same line shape is obtained. In the
former case, the effect of the interaction between impurities is stronger
and the line is wider and with some structure due to a partial splitting of
the Kondo resonance. In the second case, the result is very similar to the
sum of the spectra at both foci when only one impurity is present. This is
the result expected for weak interactions.

\section{Summary and discussion}

Using a simple impurity Anderson model in which the particular structure of
the surface states inside a corral is taken into account appropriately, the
basic physics of the mirage experiments in quantum corrals \cite{man,man2}
can be understood. The resulting space and voltage dependence of the
differential conductance $dI/dV$ is in good agreement with experiment. The
voltage dependence of $dI/dV$ observed for one Co atom on a clean Cu(111)
surface and for a Co atom at the focus of an elliptical corral built on that
surface can both be explained with the same set of parameters (see Fig. 11).

While the space dependence of $dI/dV$ is mainly determined by
non-interacting conduction electron Green's functions, the calculation of
the dependence of $dI/dV$ on bias voltage is a non trivial many-body
problem in which the particular structure of the conduction electrons for
surface states introduces additional complications. Single-particle
scattering theories were successful in explaining the space dependence of $%
dI/dV$ \cite{fiete,agam,fie}, but the voltage dependence is actually used to
adjust a phenomenological energy dependent phase shift. Therefore the
differences in the line shape as a function of the applied voltage in
different structures (like the above mentioned between an impurity on a
clean surface or inside the corral) cannot by accounted for. In contrast,
the many-body treatment is very difficult to implement for open structures,
while one-body scattering theory assuming simple interactions, allows not
only to calculate but also to optimize open structure to obtain multiple
mirages or other desired effects \cite{correa2}. On the other hand, our
approach leads to very good agreement with the space dependence of $dI/dV$
(see Fig. 9) except perhaps for the finest details which we did not attempt
to fit. In addition, it allows to understand the basic observed features,
including the mirage effects and its intensity, in terms of the interference
of wave functions in the corral (see section 7).

Experiments in which the change in differential conductance after addition
of one impurity in a quantum corral $\Delta dI/dV$ is measured should be
able to discern the relative importance of surface and bulk states in the
formation of the Kondo singlet. Measurements in circular corrals, easier to
handle theoretically, would be useful. In section 8 we presented some
results for this case. In
addition, the change of $\Delta dI/dV$ when more than one impurity is
present in the corral is very sensitive to the surface-impurity
hybridization. The observation of a quantum mirage establishes a lower bound
for this hybridization which we have estimated.

The calculations presented here are for $T=0$. Within perturbation theory,
the results can be extended easily to $T\neq 0$. Some results were presented
in Ref. \cite{lob1}. Important changes occur in the scale of the Kondo
temperature, but the behavior is similar to that already known for the
simple case explained in section 3.

An improvement of the many-body theory requires a better knowledge of 
the hybridization of the impurity with surface and bulk states
and their wave vector dependence (which we have neglected). The wave vector
dependence is expected from a jellium model \cite{pli}. In particular for
the states near the Fermi energy $\epsilon _{F}$ and wave functions decaying
as $\exp (-\kappa z)$ out of the surface, one expects $k_{||}^{2}-$ $\kappa
^{2}$ constant and therefore a smaller wave vector parallel to the surface $%
k_{||}$ implies a weaker decay rate $\kappa $ in the perpendicular $z$
direction, and therefore a larger hybridization with impurity and tip. For
surface states near $\epsilon _{F}$, $k_{||}$ is small and one would expect
a weaker decay for surface than bulk states. This suggests a stronger
relative hybridization of bulk states with the impurity in comparison with
the tip, because the former is closer to the surface. This would be
consistent with the fact that the hybridization of the tip with surface and
bulk states is of the same order \cite{knorr,burg,jean}, and the proposal
that bulk states dominate the hybridization with the impurity \cite
{knorr,schn,cor2,andrea}. However, more detailed calculations of the matrix
elements found oscillations of the matrix elements with $z$, and a stronger
relevance of surface states in the formation of the Kondo state \cite{lin}
and in the distance dependence of the observed $dI/dV$ for magnetic
impurities on clean (111) surfaces \cite{lin,meri}.

We have shown that the width of the surface conduction electrons $\delta $
plays a crucial role in the many-body theory (see section 6). We have
calculated this width for a circular confinement potential and found that it
increases linearly with energy \cite{lob1}. However, we assumed that for a
clean surface the surface states are well defined for all wave vectors and
this is not the case for energies above the Fermi energy \cite{hulb}. Since
localization involves a participation of all wave vectors, there is a
contribution to the width brought by the states of larger wave vector which
we have neglected. However, since most of the physics depends on the value
of $\delta $ at the Fermi energy and we took it as a parameter, our
conclusions are not affected.

Correlation functions of impurities inside a quantum corral have been
studied previously \cite{wil,hal,lob1,correa,correa2}. For perfect
confinement a strong enhancement should occur. However, for realistic
broadening of the surface conduction states, and distances of the order of
several nm involved in the mirage experiments, we expect that the single ion
physics dominate the RKKY interactions, and no significant magnetic
correlations are present \cite{lob1}.

\section*{Acknowledgments}

A.A.A. wants to thank Mar\'{\i}a Andrea Barral for helpful discussions. We
are partially supported by CONICET. This work was sponsored by PICT 03-12742
of ANPCyT.


\begin{references}
\bibitem{note3}  The description of the Green's function of the surface
conduction eigenstates as a sum of resonances can be demonstrated under
quite general assumptions \cite{cal} and has been worked out in detail for
the circular corral \cite{lob1,lob2}.
\bibitem{note}  In the ordinary Anderson model, for small enough
hybridization $V$ one has $T_{K}\sim \Delta \exp [-1/(\rho J_{K})]$ , where $%
\Delta =\pi \rho V^{2}$, $\rho $ is the density of states per spin at the
Fermi level, and $J_{K}=V^{2}[1/(\epsilon _{F}-E_{d})+1/(U+E_{d}-\epsilon
_{F})]$ is the Kondo coupling \cite{andrei,tsve,integ}.
\bibitem{note5}  The bulk Green's function $G_{b}^{0}(R_{1},R_{2},\omega )$
contributes a uniform background to $dI/dV$ [Eq. (\ref{didv})] through the
second term of the first Eq. (\ref{gh}). In addition, it has a contribution
to $\Delta dI/dV$ through the second Eq. (\ref{gh}) which is important only
when tip and impurity are close to each other because $G_{b}^{0}\sim
1/|R_{1}-R_{2}|^{2}$ at distances larger than a few nm \cite{pli,meri}.
\bibitem{note6} The decrease (increase) in $U$ for fixed $E_d$ leads to a 
smaller (larger) $V_s$ and $V_b$ to keep the experimentally observed width
of the Fano antiresonance, but has no other important consequences.
\bibitem{note2}  If the Fermi level falls between two energies of the
conduction states, a peak is present at $\epsilon _{F}$ for $d\gg T_{K}$,
but in any case, the structure of $\rho _{d\sigma }(\omega )$ does not show
the features that correspond to the continuous conduction density of states 
\cite{thi}.
\bibitem{note4}  The parameters were modified so that the contribution of $%
V_{b}$ and $V_{s}$ to the width of the Fano antiresonance for $\delta =40$
meV is the same.
\end{references}
\end{document}